\documentclass{aa}  

\usepackage{txfonts}
\usepackage{scalefnt}
\usepackage{pdflscape}
\usepackage{enumerate}
\usepackage{float}
\usepackage{graphicx}
\usepackage{subfig}
\usepackage{txfonts}
\begin{document} 

\newbox\grsign \setbox\grsign=\hbox{$>$} \newdimen\grdimen \grdimen=\ht\grsign
\newbox\simlessbox \newbox\simgreatbox
\setbox\simgreatbox=\hbox{\raise.5ex\hbox{$>$}\llap
     {\lower.5ex\hbox{$\sim$}}}\ht1=\grdimen\dp1=0pt
\setbox\simlessbox=\hbox{\raise.5ex\hbox{$<$}\llap
     {\lower.5ex\hbox{$\sim$}}}\ht2=\grdimen\dp2=0pt
\def\simgreat{\mathrel{\copy\simgreatbox}}
\def\simless{\mathrel{\copy\simlessbox}}
\newbox\simppropto
\setbox\simppropto=\hbox{\raise.5ex\hbox{$\sim$}\llap
     {\lower.5ex\hbox{$\propto$}}}\ht2=\grdimen\dp2=0pt
\def\simpropto{\mathrel{\copy\simppropto}}

\title{Cobalt and copper abundances in 56 Galactic bulge red giants
\thanks{Observations collected at the European  Southern  Observatory,
  Paranal,  Chile  (ESO programmes  71.B-0617A, 73.B0074A, and GTO 71.B-0196)} }
\author{
H. Ernandes\inst{1,2,3}
\and
B. Barbuy\inst{1}
\and
A. Fria\c ca\inst{1}
\and
V. Hill\inst {4}
\and
M. Zoccali\inst{5,6}
\and
D. Minniti\inst{6,7,8}
\and
A. Renzini\inst{9}
\and
S. Ortolani\inst{10,11}
}
\offprints{H. Ernandes}
\institute{
Universidade de S\~ao Paulo, IAG, Rua do Mat\~ao 1226,
Cidade Universit\'aria, S\~ao Paulo 05508-900, Brazil
\and
UK Astronomy Technology Centre, Royal Observatory, Blackford Hill, Edinburgh, EH9 3HJ, UK
\and
IfA, University of Edinburgh, Royal Observatory, Blackford Hill, Edinburgh, EH9 3HJ, UK
\and
Universit\'e de Sophia-Antipolis,
 Observatoire de la C\^ote d'Azur, CNRS UMR 6202, BP4229, 06304 Nice Cedex 4, France
 \and
 Universidad Catolica de Chile, Departamento de Astronomia y Astrofisica, Casilla 306, Santiago 22, Chile
 \and
Millennium Institute of Astrophysics, Av. Vicuna Mackenna 4860, 782-0436, Santiago, Chile
\and
Departamento de Ciencias Fisicas, Facultad de Ciencias Exactas, Universidad Andres Bello, Av. Fernandez Concha 700, Las Condes, Santiago, Chile
\and
Vatican Observatory, V00120 Vatican City State, Italy
\and
Osservatorio Astronomico di Padova, Vicolo
 dell'Osservatorio 5, I-35122 Padova, Italy
\and
Universit\`a di Padova, Dipartimento di Fisica e Astronomia, Vicolo
 dell'Osservatorio 2, I-35122 Padova, Italy
\and
INAF-Osservatorio Astronomico di Padova, Vicolo dell'Osservatorio 5,
I-35122 Padova, Italy
}

   \date{}

 
  \abstract
      { The Milky Way bulge is an important tracer of the early formation
          and chemical enrichment of the Galaxy. 
          The abundances of different iron-peak elements in field bulge stars
          can give 
        information on the nucleosynthesis processes that took place in the
        earliest supernovae. Cobalt (Z=27) and copper (Z=29) are particularly interesting. }
      {We aim to identify the nucleosynthesis processes responsible for
      the formation of the iron-peak elements Co and Cu.}
   {We  derived abundances of
the iron-peak elements cobalt and copper
in 56 bulge giants, 13 of which were red clump stars.
High-resolution spectra were obtained using  FLAMES-UVES at the ESO Very 
Large Telescope by our group in 2000-2002, which appears to be the highest quality  sample
of high-resolution data on bulge red giants obtained in the literature to date.
   Over the years we  have derived the abundances of C, N, O, Na, Al, Mg; the iron-group
   elements Mn and Zn; and neutron-capture elements. In the present work we
   derive abundances of the iron-peak elements cobalt and copper. We also compute
   chemodynamical evolution models to interpret the observed behaviour of these elements as a
   function of iron.}
   {The sample stars show mean values of [Co/Fe]$\sim$0.0 at all metallicities, and
   [Cu/Fe]$\sim$0.0 for [Fe/H]$\geq$-0.8 and decreasing towards lower metallicities
   with a behaviour of a secondary element. }
   {We conclude that [Co/Fe] varies in lockstep with [Fe/H], which indicates that it should
   be produced in the alpha-rich freezeout mechanism in massive stars. Instead  [Cu/Fe] follows the behaviour
   of a secondary element towards lower metallicities, indicating its production in the
   weak s-process nucleosynthesis in He-burning and later stages. The chemodynamical models presented here confirm the behaviour of these two elements (i.e. [Co/Fe] vs. [Fe/H]$\sim$constant and [Cu/Fe] decreasing with
   decreasing metallicities).}


   \maketitle
%

\section{Introduction} 

The detailed study of element abundances in the Milky Way bulge can  inform on the chemical enrichment processes in the Galaxy, and on the early stages of the Galaxy formation.  Field stars in the Galactic bulge are
  old (Renzini et al. 2018, and references therein), and bulge globular
  clusters, in particular the moderately metal-poor ones, are very old
  (e.g. Kerber et al. 2018, 2019, Oliveira et al. 2020). The study of bulge
  stars can therefore provide hints on the chemical enrichment of the earliest
stellar populations in the Galaxy.
  Abundance ratio indicators have been extensively
   used in the literature and interpreted in terms
   of nucleosynthesis typical of different types of supernovae
   and chemical evolution models. The studies are most usually based on the
   alpha-elements O, Mg, Ca, and Si, and on Al and Ti, which behave like
   alpha-elements that are enhanced in metal-poor stars (e.g.
   Mishenina et al. 2002, Cayrel et al.
   2004, Lai et al. 2008), in the Galactic bulge (e.g. McWilliam 2016, Fria\c ca \& Barbuy 2017),
   and elliptical galaxies (e.g. Matteucci   \& Brocato 1990).
   The alpha-element enhancement in old stars
   is due to a fast chemical enrichment by supernovae type II (SNII).
   Other independent indicators have so far been less well studied, notably
   iron-peak elements, s-elements, and r-elements. Ting et
   al. (2012) aimed to identify which groups of elements are
   independent indicators of the supernova type that produced them.
   Their study reveals two types of SNII: one that produces mainly
   $\alpha$-elements and one that produces both $\alpha$-elements
   and Fe-peak elements with a large enhancement of heavy Fe-peak
   elements, which may be the contribution from hypernovae. This shows
   the importance of deriving Fe-peak element abundances.

   The Fe-peak elements have atomic numbers in the range 21 $\leq$ Z $\leq$ 32.
   The lower iron group includes Sc, Ti, V, Cr, Mn, and Fe and the upper
   iron-group contains Co, Ni, Cu, and Zn, and probably also Ga and Ge.
   Most of these elements vary in lockstep with Fe   as a function of metallicity, with the exception of
   Sc, Ti, Mn, Cu, and Zn, and perhaps Co (e.g.
   Gratton 1989, Nissen et al. 2000, Sneden et al. 1991, Cayrel et al. 2004,
   Ishigaki et al. 2013, da Silveira et al. 2018). This occurs
   because most Fe is produced in SNIa,  whereas some of the iron-peak
   elements, in particular Co and Cu, are well produced in massive stars
   (Sukhbold et al. 2016, and references therein).

   We previously analysed the iron-peak elements Mn and Zn
   (Barbuy et al. 2013, 2015, da Silveira et al. 2018) in the same sample of field stars
studied in the present work, as well as Sc, V, Cu, Mn, and Zn in bulge globular cluster
stars (Ernandes et al. 2018). In this paper we analyse abundances
of the iron-peak elements cobalt and copper.
These two elements, and copper in particular,
deserve attention because the nucleosynthesis processes
that produce them have been discussed over the years in
the literature. The production of Cu in
massive stars as a secondary product was only  challenged by Mishenina
et al. (2002), who argued that a sum of a secondary and a primary process
would be needed to explain the behaviour of [Cu/Fe] versus [Fe/H] in metal-poor
stars. 
Bisterzo et al. (2004) concluded that most Cu derives from a secondary
weak-s process in massive stars; a small primary contribution of $\sim$5\%
in the Sun would be due to the decay of $^{63,65}$Zn, and this becomes
dominant for [Fe/H]$<$-2.0. On the other hand,
asymptotic giant branch (AGB) stars and SNIa contribute little to Cu. 
Pignatari et al. (2010) presented nucleosynthesis calculations showing
an increased production of Cu from a weak-s process in massive stars.
Romano \& Matteucci (2007) concluded that Cu enrichment is due to
a primary contribution from explosive nucleosynthesis in SNII, and
a weak s-process in massive stars.
Lai et al. (2008) data on halo stars agreed with these models.

  According to Woosley \& Weaver (1995, hereafter WW95),
Limongi et al. (2003), and Woosley et al. (2002),
the upper iron-group elements are mainly synthesized in  two processes: either neutron capture on iron-group nuclei during He burning and later burning stages (also called the weak {\it s}-component);  or the $\alpha$-rich freezeout in the deepest layers. Both cobalt and copper are produced as primary elements in the $\alpha$-rich freezeout   and as secondary elements in the weak {\it s}-process in massive stars.
The relative efficiency of these two contributions to the nucleosynthesis of Co and Cu can be tested
by deriving their abundances in the Galaxy. Abundances gathered so far, in the Galactic bulge in particular,  indicate that copper behaves as a secondary element, therefore with a significant contribution from the weak {\it s}-process.  Cobalt, which appears to vary in lockstep with Fe,   seems instead to be mostly contributed from the $\alpha$-rich freezeout mechanism (Barbuy et al. 2018, Woosley, private communication).

Very few previous analyses of iron-peak elements in Galactic bulge stars
are available in the literature.
For copper, Johnson et al. (2014) and Xu et al. (2019) are so far the only
available  data derived from moderately high-resolution spectra.
 For cobalt,  Johnson et al. (2014) present results from moderately high-resolution spectra, Schultheis et al. (2017) from near-infrared (NIR) spectra,
 and Lomaeva et al. (2019) from high-resolution spectra.
 
Our observations are outlined in Sect. 2. In Sect. 3 
 we list the basic stellar parameters, report atomic constants under study for the lines of Co and Cu, and describe the
 abundance derivation of Co and Cu.
 Chemical evolution models
are presented in Sect. 4. The results
are discussed in Sect. 5, and conclusions are drawn in Sect. 6.


\section{Observations}

The present data consist of high-resolution spectra 
of 43 bulge red giants,
chosen to have one magnitude brighter than the red clump,
from ESO programmes  71.B-0617A and 73.B0074A (PI: A. Renzini)
obtained with the FLAMES-UVES spectrograph  (Dekker et al. 2000)
 at the 8.2 m Kueyen Very Large Telescope at the Paranal Observatory of the
European Southern Observatory (ESO).
The stars were observed in four fields, namely 
Baade's Window (BW) (l=1.14$^{\circ}$, b=-4.2$^{\circ}$), 
a field at $\rm b=-6^{\circ}$ (l=0.2$^{\circ}$, b=$-6^{\circ}$),
the Blanco field (l=0$^{\circ}$, b=$-12^{\circ}$), 
and a field near NGC 6553 (l=5.2$^{\circ}$, b=$-3^{\circ}$).
Thirteen additional red clump bulge giants were observed in programme
GTO 71.B-0196 (PI: V. Hill), as described in Hill et al. (2011).

The mean wavelength coverage is 4800-6800 {\rm \AA}.
With the UVES (Ultraviolet and Visual Echelle Spectrograph)
standard setup 580, the resolution is R $\sim$ 45 000 for a 1
arcsec slit width, given that the fibres are 1.0" wide.
Typical signal-to-noise ratios obtained by considering average values at
different wavelengths vary in the range 30 $\leq$ S/N $\leq$ 280 per pixel in the
programme stars. Here the analysis is based uniquely on the UVES spectra,
but it is noteworthy that
the same sample of stars was also observed with the GIRAFFE
spectrograph, as part of a larger sample (Zoccali et al. 2008), with the
purpose of validating their abundance analysis at the lower resolution
(R $\sim$ 22,000) of GIRAFFE.
 
As described in Zoccali et al. (2006), Lecureur et al. (2007), and
Hill et al. (2011), the spectra were reduced using the 
FLAMES-UVES pipeline, 
including bias and inter-order background subtraction, flat-field correction,
extraction, and wavelength calibration (Ballester et al. 2000).

This sample of stars had the abundances of O, Na, Mg, and Al studied
in Zoccali et al. (2006) and Lecureur et al. (2007). 
The C, N, and O abundances were revised in Fria\c ca \& Barbuy (2017).
The iron-peak elements Mn and Zn were studied
in Barbuy et al. (2013, 2015) and da Silveira et al. (2018),
and heavy elements in van der Swaelmen et al. (2016).
 In summary,
the abundances of C, N, O, Na, Mg, Al, Mn, Zn, and heavy elements
were derived.
Gonz\'alez et al. (2011) derived abundances of Mg, Si, Ca, and Ti for
 a GIRAFFE counterpart of the sample, obtained at R $\sim$ 22 000). 
 Da Silveira et al. (2018)
 derived O and Zn from GIRAFFE data in two fields.

It is interesting to note that this data set, including both the
high-resolution UVES data as well as the moderately high-resolution 
GIRAFFE data, has become an important reference for bulge studies; from this same ESO programme, Johnson et al. (2014) analysed GIRAFFE data
for 156 red giants in the Blanco and near-NGC 6553 fields, 
and Xu et al. (2019) reanalysed 129 of these same stars.
J\"onsson et al. (2017) reanalysed UVES spectra of a sub-sample
of 33 stars from our sample of 43 red giants,
and additionally  analysed two other stars, BW-b1 and B2-b8, 
that were observed but not included
in the studies by Zoccali et al. (2006, 2008).
A comparison of stellar parameters between Zoccali et al. (2006) and
Lecureur et al. (2007) relative to J\"onsson et al. (2017)
is discussed in da Silveira et al. (2018). 
The same sub-sample that was reanalysed by J\"onsson et al. (2017) was further
analysed by Forsberg et al. (2019), Lomaeva et al. (2019), and 
Grisoni et al. (2020) for different elements, adopting their
own stellar parameters.
Finally, Schultheis et al. (2017) compared
APOGEE (Apache Point Observatory Galactic Evolution
Experiment) results  with stars in common with
Zoccali et al. (2008)'s results for stars  observed with GIRAFFE.


\section{Abundance analysis}

The stellar parameters 
effective temperature (T$_{\rm eff}$), gravity (log g), 
metallicity ([Fe/H]\footnote{Here we adopted the usual
spectroscopic notation, that [A/B] = log(N$_{\rm A}$/N$_{\rm B}$)$_{\star}$ $-$
log(N$_{\rm A}$/N$_{\rm B}$)$_{\odot}$ and $\epsilon$(A) = log(N$_{\rm
A}$/N$_{\rm B}$) + 12 for both elements A and B.}), and microturbulence
velocity (v$_{\rm t}$) are adopted from our previous determinations
(Zoccali et al. 2006, 2008, Lecureur et al. 2007), which we summarize below.

 The VIJKH de-reddened magnitudes  were combined
to obtain photometric temperatures from V-I, V-J, V-H, and V-K colours.
The mean of the four values
was used as a first guess for a spectroscopic analysis. Photometric
gravity was calculated from the classical relation

\[
\log g_*=4.44+4\log \frac{T_*}{T_{\odot}}+0.4(M_{\rm bol}-4.75)+\log \frac{M_*}{M_{\odot}}, 
\]

\noindent
 adopting T$_{\odot}$=5770 K, M$_*$=0.85 M$_{\odot}$, M$_{\rm bol \odot}$
= 4.75, and a mean distance of 8 kpc for the Galactic bulge.

The equivalent widths for selected lines of Fe, Na, Mg, Al,
Si, Ca, Sc, Ti, and Ni were measured using the 
code DAOSPEC (Stetson and Pancino 2008). The selection
of clean Fe lines and their atomic parameters was compiled
using a spectrum of $\mu$ Leo as reference (Lecureur et al. 2007).
 
The LTE abundance analysis was performed using an updated
version of the code ABON2 (Spite 1967) and  MARCS models
(Gustafsson et al. 2008).
 Excitation equilibrium was imposed on the \ion{Fe}{I} lines
in order to refine the photometric T$_{\rm eff}$, while photometric 
gravity was imposed even if ionization equilibrium was not fulfilled.

Elemental abundances were obtained through line-by-line spectrum synthesis
calculations. The calculations of synthetic spectra were carried out 
using the PFANT code described in Barbuy et al. (2018b), where
molecular lines of the  CN  A$^2$$\Pi$-X$^2$$\Sigma$, C$_2$  Swan 
A$^3$$\Pi$-X$^3$$\Pi$ and TiO A$^3$$\Phi$-X$^3$$\Delta$ $\gamma,$ and
B$^3$$\Pi$-X$^3$$\Delta$ $\gamma$' systems are taken into account.
The MARCS model atmospheres are adopted (Gustafsson et al. 2008).

The abundances derived line-by-line are reported in Table
\ref{atmos3}. The final mean abundances are given in
the last three columns of  Table \ref{atmos2}, where the final mean values of [Cu/Fe] and [Co/Fe] in LTE and NLTE-corrected are reported.

\begin{figure*}
    \centering
    \includegraphics[width=1.5\columnwidth]{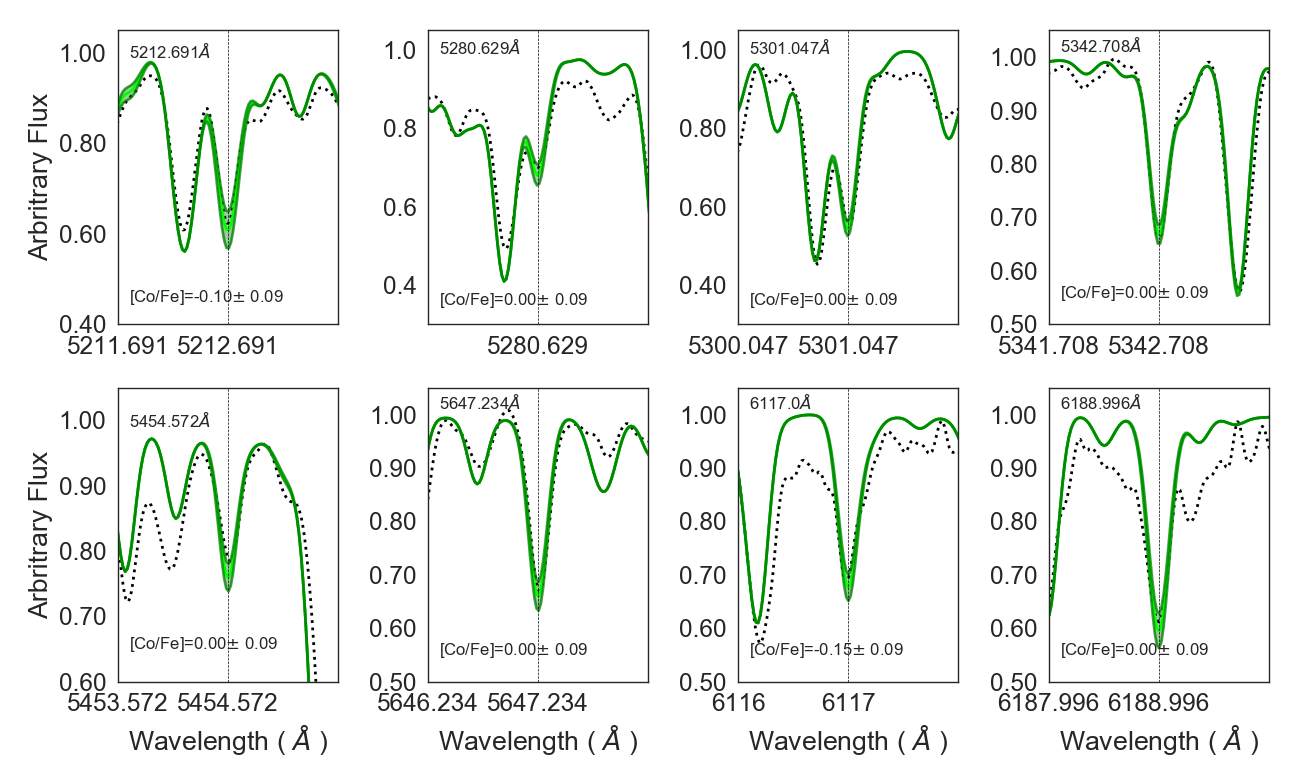}
    \caption{Fits of synthetic spectra to the eight observed lines
      of \ion{Co}{I} in star BWc-4. The dotted line is the observed spectrum.
      The green lines correspond to the value adopted, and with
      [Co/Fe]=+0.09 and -0.09. }
    \label{fitbwc4co}
\end{figure*}

\begin{figure}
    \centering
    \includegraphics[width=0.9\columnwidth]{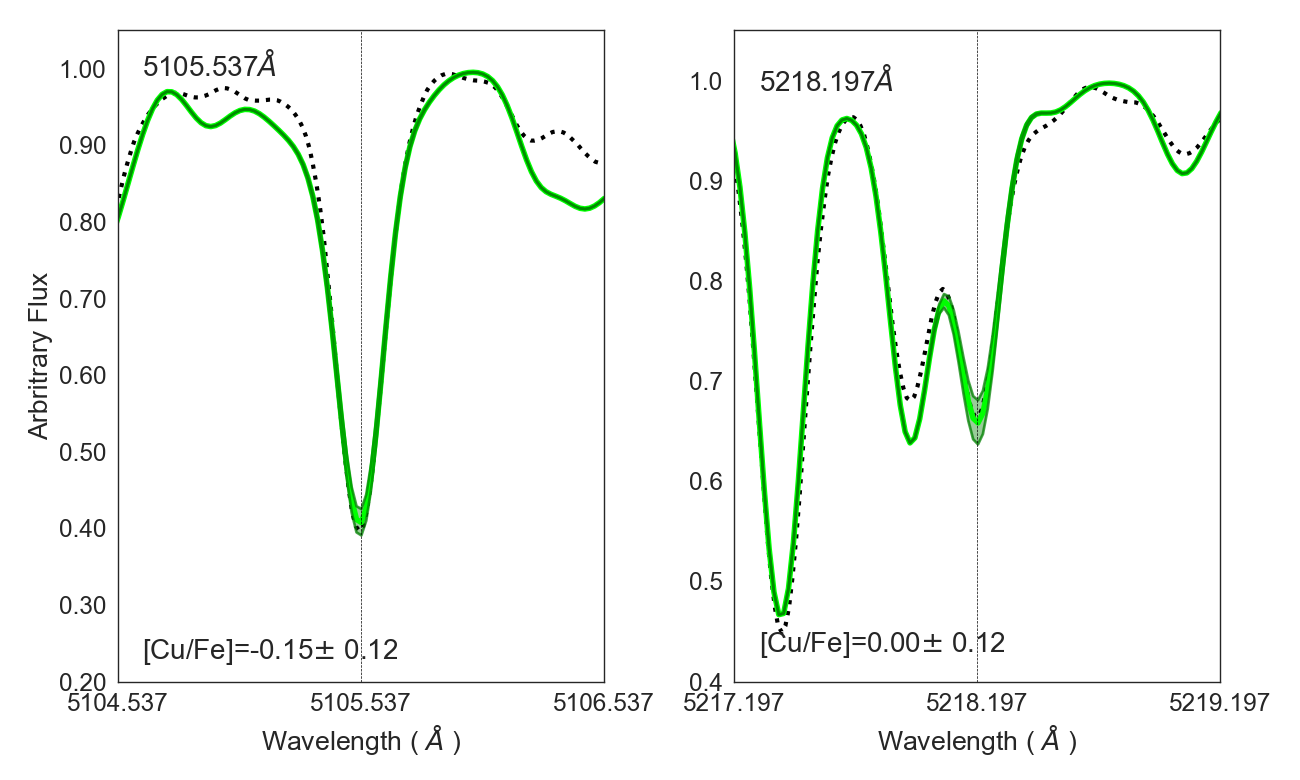}
    \caption{Fits of synthetic spectra to the two observed lines
      of \ion{Cu}{I} in star BL-7. The dotted line is the observed spectrum.
      The green lines correspond to the value adopted, and with [Cu/Fe]=+0.12 and -0.12.}
    \label{fitbl7cu}
\end{figure}

Figure \ref{fitbwc4co} shows
the fit to the eight \ion{Co}{I} lines in star BWc-4. 
Figure \ref{fitbl7cu}  shows the fit to the \ion{Cu}{I} 5105.537
and 5218.197 {\rm \AA} lines for star BL-7.

\subsection{Line parameters: hyperfine structure, oscillator strengths, and solar
abundances}

We derive cobalt and copper abundances for the 56 sample stars
using the lines of \ion{Co}{I}  and \ion{Cu}{I} reported in Table \ref{lines2}.
The oscillator strengths and the hyperfine  structure (HFS)
we adopted are described below.


{\it Cobalt: \ion{Co}{I} lines} 

Cobalt has the unique species $^{59}$Co (Asplund et al. 2009).
The HFS was taken into account by applying the code made available 
by McWilliam et al. (2013) together with the A and B constants
reported in Table \ref{coculines} that were adopted
from Pickering et al. (1996). Cobalt has a nuclear spin I = 7/2.
Central wavelengths and excitation potential values from
Kur\'ucz (1993)\footnote{http://kurucz.harvard.edu/atoms.html}, the oscillator strengths 
from Kur\'ucz (1993),
NIST\footnote{http://physics.nist.gov/PhysRefData/ASD/lines$_-$form.html}
(Martin et al. 2002), 
 and VALD (Piskunov et al. 1995), and the final values adopted are presented in Table \ref{lines2}.

Tables \ref{hfsCo1}, \ref{hfsCo2}, and \ref{hfsCo3}
 show the HFS components of \ion{the  Co}{I} lines studied.
All these lines were checked by comparing synthetic spectra to
high-resolution spectra of the Sun  (using the same instrument settings as the present sample of spectra \footnote{http://www.eso.org/\-observing/\-dfo/\-quality/\-UVES/\-pipeline/solar{$_{-}$}spectrum.html}), Arcturus (Hinkle et al. 2000), and the metal-rich giant star 
$\mu$ Leo (Lecureur et al. 2007).
We adopted the following stellar parameters: effective temperature (T$_{\rm eff}$), surface gravity (log~g), 
metallicity ([Fe/H]), and microturbulent velocity (v$_{\rm t}$)  of
 (4275 K, 1.55, -0.54, 1.65 km.s$^{-1}$) for Arcturus from 
Mel\'endez et al. (2003), and (4540 K, 2.3, +0.30, 1.3 km.s$^{-1}$) for $\mu$ Leo from Lecureur et al. (2007).
The adopted abundances for the Sun, 
Arcturus, and $\mu$ Leo are reported in Table \ref{adopted_ab}. 

{\it Copper: \ion{Cu}{I} lines}

Copper abundances were derived from the two \ion{Cu}{I} lines at 5105 and 5218 {\rm \AA} already employed and described in detail in
Ernandes et al. (2018). 
The 5782 {\rm \AA} line is not available in the UVES spectra. Isotopic fractions
of  0.6894 for $^{63}$Cu and 0.3106 for $^{65}$Cu (Asplund et al. 2009),
as well as the HFS structure
 as given in Ernandes et al. (2018), are adopted.

Figures \ref{cobaltlines} and \ref{copperlines} show the fits to the spectra
of the Sun, Arcturus, and $\mu$ Leo. The \ion{Cu}{I} atomic parameters and 
fits to these reference stars were already extensively discussed in
Ernandes et al. (2018).
 
\begin{table*}
\begin{flushleft}
\centering
\caption{Central wavelengths and total oscillator strengths.}             
\label{lines2}      
\begin{tabular}{lcccccccccccccc}     
\noalign{\smallskip}
\hline\hline    
\noalign{\smallskip}
\noalign{\vskip 0.1cm} 
species & {\rm $\lambda$} ({\rm \AA}) & {\rm $\chi_{ex}$ (eV)} & {\rm gf$_{Kurucz}$} &
 {\rm gf$_{NIST}$} & {\rm gf$_{VALD}$}  & {\rm gf$_{adopted}$}  \\
\noalign{\vskip 0.1cm}
\noalign{\hrule\vskip 0.1cm}
\noalign{\vskip 0.1cm}
CoI    & 4749.669 & 3.053457        & -0.321  &---   &-0.236   &-0.321   \\
CoI    & 5212.691 & 3.514439        & -0.110  &-0.11 &-0.110   &-0.110   \\
CoI    & 5280.629 & 3.628984        & -0.030  &-0.03 &-0.030   &-0.030   \\
CoI    & 5301.039 & 1.710426        & -2.000  &-1.99 &-2.000   & -2.000  \\
CoI    & 5342.695 & 4.020881        &  0.690  &---   & 0.741   & 0.690   \\
CoI    & 5454.572 & 4.071888        & -0.238  &---   &+0.238   & +0.238   \\
CoI    & 5647.234 & 2.280016        & -1.560  &-1.56 &-1.560   &-1.560   \\
CoI    & 6117.000 & 1.785283        & -2.490  &-2.49 &-2.490   &-2.490   \\
CoI    & 6188.996 & 1.710426        & -2.450  &-2.46 &-2.450   & -2.450   \\
CuI    & 5105.537 & 1.389035 & -1.516 & -1.50 & hfs   & -1.52 \\ 
CuI    & 5218.197 & 3.816948 & 0.476  & 0.264 & hfs   & +0.124  \\
\noalign{\vskip 0.1cm}
\noalign{\hrule\vskip 0.1cm}
\noalign{\vskip 0.1cm}  
\hline                  
\end{tabular}
\end{flushleft}
\end{table*}

\begin{table}
\caption{Adopted abundances for the Sun, Arcturus, and $\mu$Leo.
References:
[1]: Grevesse et al. (1996); Grevesse \& Sauval (1998); Asplund et al. (2009); Lodders (2009);
[2]: Steffen et al. (2015); 
[3]: Ram\'irez \& Allende Prieto (2011); 
[4]: Mel\'endez et al. (2003)
[5]: Barbuy et al. (2015); Fria\c ca \& Barbuy (2017);
[6]: Smith et al. (2013); 
[7]: Gratton \& Sneden (1990);
[8]: Lecureur et al. (2007)
[9]: Barbuy et al. (2014)
[10]: Smith \& Ruck (2000)
[11]: Scott et al. (2015a,b)
[12]: McWilliam et al. (2013)}   
\label{adopted_ab} 
\centering                  
\begin{tabular}{l c c c c} 
\hline\hline             
El. & Z & A(X)$_{\odot}$ & A(X)$_{Arcturus}$ & A(X)$_{\mu Leo}$ \\ 
\hline  
Fe & 26 & 7.50 [1] & 6.96 [4]  & 7.80 [5] \\                
C  & 6  & 8.55 [1] & 7.79 [8] & 8.55 [5] \\   
N  & 7  & 7.97 [1] & 7.65 [8] & 8.83 [5]\\   
O  & 8  & 8.77 [2] & 8.62 [9] & 8.97 [5]\\   
Na & 11 & 6.33 [1] & 5.90 [3] & 7.07 [10]\\   
Mg & 12 & 7.58 [1] & 7.41 [3] & 7.91 [10]\\   
Al & 13 & 6.47 [1] & 6.26 [3] & 6.90 [6]\\   
Si & 14 & 7.55 [1] & 7.34 [11] & 8.02 [7]\\   
K  & 19 & 5.12 [1] & 4.99 [3] & 5.63 [6]\\   
Ca & 20 & 6.36 [1] & 5.94 [3] & 6.62 [6]\\   
Sc & 21 & 3.17 [1] & 2.86 [5] & 3.34 [7]\\   
Ti & 22 & 5.02 [1] & 4.74 [13] & 5.39 [10]\\   
V  & 23 & 4.00 [1] & 3.58 [3] & 4.34 [7]\\   
Cr & 24 & 5.67 [1] & 5.08 [3] & 5.97 [7]\\   
Mn & 25 &  5.39 [1] & 4.71 [12] & 5.70 [7]\\  
Co & 27 & 4.92  [1]  & 5.11 [3]  & 4.93 [7] \\
Ni & 28 & 6.25 [1] & 5.77 [3] &  6.60 [10]\\
Cu & 29 & 4.21  [1]  & 4.09 [3]  & 4.46 [10]\\
Zn & 30 & 4.60 [1] & 4.06 [5] & 4.80 [5]\\   
\hline                          
\end{tabular}
\end{table}


\subsection{Non-local thermodynamic equilibrium corrections\label{NLTE}}
We applied the NLTE corrections for each cobalt line following the same method used by 
Kirby et al. (2018), with the formalism of Bergemann \& Cescutti (2010) and Bergemann et al. (2010) \footnote{http:\/\/nlte.mpia.de\/gui-siuAC\_secE.php}.
The derivation of corrections from the online code made
available requires the choice of atmospheric model, inclusion of
 stellar parameters of each star, and the line list, followed by the atomic number (Z) under study.
The corrections so derived line-by-line for the Co abundances are reported in Table \ref{atmos4}, and final NLTE-corrected  Co abundance values are given in Table \ref{atmos2}.

\subsection{Uncertainties\label{uncertainties}}
As with our previous papers regarding this sample of spectra,
and given that the final adopted atmospheric parameters for
the program stars were based on Fe I and Fe II lines together with photometric gravities, we have adopted their
estimated uncertainties in the atmospheric parameters (i.e.\ $\pm$ 100 K for
temperature, $\pm$ 0.20 for surface gravity, and $\pm$ 0.20 kms$^{-1}$ for
microturbulent velocity). 
In Table \ref{errors}
we compute Co and Cu abundances for the metal-rich star B6-f8
and the metal-poor star BW-f8 by changing
their parameters by these amounts. The errors computed
by adopting models with $\Delta$T$_{eff}$=+100K, $\Delta$log~g=+0.2,
and $\Delta$v$_{t}$=+0.2 km.s$^{-1}$, as well as final errors, are shown
in Table \ref{errors}.


For comparison purposes, we have listed the stars that were also analysed by Johnson et al. (2014)
and J\"onsson et al. (2016) in Table 6,
reporting the respective stellar parameters they adopted.
Johnson et al. (2014)  analysed their corresponding GIRAFFE spectra, while
J\"onsson et al. (2016) reanalysed the same UVES data as
Zoccali (2006, 2008) and Lecureur et al. (2007); these data and stellar parameters are the
same as given in Lomaeva et al. (2019).

The differences in stellar parameters between the present ones
adopted from Zoccali et al. (2006, 2008), Lecureur et al. (2007),
and the reanalysis by  J\"onsson et al. (2017)
were discussed in da Silveira et al. (2018).
 As reported in Sect. 3, the present parameters (see Sect. 3) were obtained 
  by applying excitation equilibrium imposed on the \ion{Fe}{I} lines
  in order to refine the photometric T$_{eff}$, and photometric gravity
  was imposed.

  The Lomaeva et al. (2019) parameters, adopted from J\"onsson et al. (2017),
were obtained by using the software Spectroscopy Made Easy (SME;
Valenti \& Piskunov 1996). The SME software simultaneously
fits stellar parameters and/or abundances by fitting calculated
synthetic spectra to an observed spectrum.
All the stellar parameters (T$_{\rm eff}$ , log g, [Fe/H],
and v$_{\rm t}$) were derived simultaneously using relatively weak, unblended
\ion{Fe}{I}, \ion{Fe}{II} , and \ion{Ca}{I} lines
and gravity-sensitive \ion{Ca}{I}-wings.

In the mean, the differences in parameters amount to
$\Delta$T$_{\rm eff}$\-(J\"onsson+17\--\-Zoccali+06)\-=\--94 K
in effective temperatures and $\Delta$log~g\-(J\"onsson+17\--Zoccali+06)\-=\-+0.46 in gravities.
 The gravities adopted by J\"onsson et al. (2017) are possibly
 too high because the sample stars were chosen to have one magnitude
 brighter than the red clump or horizontal branch.
 It is well known that the red clump stars have rather homogeneous
 gravity values of log~g$\sim$2.2 that can go up to log~g$\sim$2.5 
 at most, depending on metallicity (Girardi 2016), and should be
 around log~g$\sim$2.3 for the stellar parameters of the present 
 metallicities. Therefore, it appears natural that red giants
 located at one magnitude
 above the red clump should have gravities around log~g$\sim$2.0
 (or lower).  On the other hand, the patchy extinction
   towards the bulge might arguably accommodate larger gravities
   for the sample stars, as assumed by J\"onsson et al. (2017).
In any case, we prefer to keep the parameters from our group
 for the sake of homogeneity of elemental abundances 
between this paper and the previous ones. Furthermore, since we have
56 stars, including 33 in common with J\"onsson et al. (2017),
it is also important to have an internal consistency in the analysis
of the 56 stars.

 A check of lines used by each author can explain some
differences in the results, as follows.
(i) Comparison of lines used for cobalt:
Johnson et al. (2014) used the \ion{Co}{I} 5647.23 and
6117.00 {\rm \AA} lines. Lomaeva et al. (2019) only used
 the UVES spectra from the red arm and
relied on the \ion{Co}{I} 6005.020, 6117.000, 6188.996, and 6632.430 {\rm \AA}. 
We have used lines from both the red arm and the blue arm spectra, as listed in Table \ref{lines2};
(ii) Comparison of lines used for copper:
Johnson et al. (2014) and Xu et al. (2019) 
used the same \ion{Cu}{I} 5782.11 {\rm \AA} line
for the same stars, which is a well-known suitable
line with identified HFS structure.

\begin{table*}
\begin{flushleft}
\scalefont{0.7}
\centering
\caption{LTE abundances of Co and Cu derived in the present work.}             
\label{atmos3}      
\centering          
\begin{tabular}{l@{}ccccccccccccc}     
\noalign{\smallskip}
\hline\hline    
\noalign{\smallskip}
\noalign{\vskip 0.1cm} 
Star & [Fe/H] &  [Cu/Fe] & [Cu/Fe] & [Co/Fe]  & [Co/Fe]   & [Co/Fe]  & [Co/Fe]  & [Co/Fe]   & [Co/Fe]  & [Co/Fe]   & [Co/Fe]  & \\ 
    &   &  5105.5374 {\rm \AA} & 5218.1974 {\rm \AA} &  5212.691  {\rm \AA}  & 5280.629 {\rm \AA}   & 5301.047 {\rm \AA}   &  5342.708 {\rm \AA}   & 5454.572 {\rm \AA}  & 5647.234 {\rm \AA}   & 6117.000 {\rm \AA}    &  6188.996 {\rm \AA}   &  \\    
\noalign{\vskip 0.1cm}
\noalign{\hrule\vskip 0.1cm}
\noalign{\vskip 0.1cm}  
\noalign{\vskip 0.1cm}
\noalign{\hrule\vskip 0.1cm}
\noalign{\vskip 0.1cm}  
 B6-b1   & 0.07   & -0.30 &-0.10 &  -0.15 &  -0.25 & -0.30 & -0.30 & -0.30 & -0.15 &  -0.15 & 0.00 \\ 
 B6-b2   & -0.01  & -0.15 &0.20 &   -0.20 &  +0.00 &  +0.00 & -0.30 &  -0.25 &  -0.25 &  +0.00 & -0.20  \\ 
 B6-b3   & 0.10   & 0.05 &      0.00 &   -0.20 & 0.00 & 0.00 &  -0.15 & -0.10 &  0.00 & 0.00 & 0.05  \\ 
 B6-b4   & -0.41  & 0.00 &      -0.15 &   -0.15 &  -0.15 &  0.00 &  -0.15 & 0.00 &  0.00 &  0.00 & 0.00  \\ 
 B6-b5   & -0.37  & 0.35 &      -0.10 &   0.00 & 0.00 &  -0.15 &  -0.15 &  0.00 &  0.00 & -0.15 &  +0.05   \\ 
 B6-b6   & 0.11   & -0.05 &0.00   &  -0.15 & 0.00 &  -0.10 &  -0.10 &  -0.15 &  -0.15 &  -0.10 &  +0.10   \\ 
 B6-b8   & 0.03   & -0.30 &0.00  &   -0.28 & 0.00 & -0.30 & -0.30 & -0.30 & -0.10 & -0.30 &  -0.05  \\

 B6-f1   & -0.01  &-0.30 &      -0.10  &  -0.15 & 0.00 & -0.15 &  -0.30 &  -0.15 &  -0.10 & -0.30 &  0.00  \\
 B6-f2   & -0.51  &-0.30 &      -0.10  &  0.00 &  0.00 & --- &  0.00 &  -0.15 &  0.00 & -0.30 &  ---   \\ 
 B6-f3   & -0.29  &0.10 &       0.00  &   0.00 & 0.00 &  0.00 &  0.00 &  -0.10 &  +0.10 &  0.00 &  +0.10  \\ 
 B6-f5   & -0.37  &-0.30 &      0.15  &    -0.10 & 0.00 &  0.00 &  -0.15 &  -0.15 &  -0.05 &  0.00 & +0.15 \\ 
 B6-f7   & -0.42  &-0.35 &      0.00   &   0.00 &  0.00 &  0.00 & 0.00 & 0.00 & 0.00 &  0.00 &  0.00  \\ 
 B6-f8   &  0.04  &0.30 &       -0.10    &  0.00 &  0.00 &  +0.10 &  0.00 &  -0.15 &  +0.05 & -0.15 &  +0.08  \\

 BW-b2   & 0.22   &---& -0.15 &   0.00 & 0.00 &  -0.15 & --- &  -0.20 &  -0.20 &  -0.25 &  0.00  \\ 

 BW-b4   & 0.07   &---& -0.30  &  -0.20 & -0.10 & -0.30 &  +0.05&  -0.30 &  -0.15  &  +0.15 & -0.15  \\
 BW-b5   & 0.17   &---& -0.35  & 0.00 & 0.00 &  -0.05 &  -0.10 &  0.00 &  0.00 &  0.00 &  0.00  \\ 
 BW-b6   & -0.25  &---& -0.30  &  0.00 &  -0.05 &  -0.15 &  -0.10 & 0.00 &  0.00 &  -0.15 & 0.00  \\
 BW-b7   & 0.10   &---& -0.25 & -0.30 & 0.00 & -0.30 &  0.00 & -0.30 &  -0.30 &  -0.15 &  -0.10 \\ 

 BW-f1   & 0.32   & -0.40 & -0.40 &   0.00 &  --- &  0.00 & -0.30 & 0.00 &  0.00 &  0.00 & 0.00  \\ 
 BW-f4   & -1.21  & -1.00 &     -0.60  &  --- & --- & --- & 0.00 &  0.00: &  0.00: &  0.00 &  0.00  \\ 
 BW-f5   & -0.59  & 0.00 &      -0.30  &  +0.05 &  0.00 & -0.30 &  0.00 & -0.15 &  0.00 &  -0.10 & 0.00  \\ 
 BW-f6   & -0.21  & ---  &      -0.50  & 0.00 &  0.00 & - 0.10 &  -0.15 &  -0.05 & 0.00& 0.00 &  +0.30 \\ 
 BW-f7   & 0.11   &    ---   &  ---  &  -0.12 &  0.00 &  -0.30 & --- & -0.30 & -0.30 & -0.25 &  0.00  \\ 
 BW-f8   & -1.27  & -0.70 &     -0.60  &  0.00: & --- & --- & 0.00 &  +0.30 & 0.00 &  -0.10: & ---  \\ 
 BL-1    & -0.16  & 0.00&       0.10 &   0.00 &  +0.15 &  +0.35 &  +0.30 &  0.00 &  +0.30 &  -0.15 &  0.00   \\ 
 BL-3    & -0.03  & 0.10&       -0.30 & 0.00 &  0.00 &  0.00 &  +0.05 &  0.00 &  +0.05 &  -0.25 &  0.00 \\ 
 BL-4    & 0.13   & 0.30&       0.15 & 0.00 & 0.00 & +0.10 & -0.10 & -0.10 & +0.12 &  -0.15 &  +0.12 \\ 
 BL-5    & 0.16   & 0.00&       0.00 & 0.00 &  0.00 & 0.00 &  -0.15 &  -0.15 & -0.15 &  -0.15 &  0.00 \\ 
 BL-7    & -0.47  & 0.15&       0.00 &  +0.05 &  +0.10 &  +0.10 &  +0.15 &  +0.10 &  +0.10 &  -0.10 &  -0.05  \\ 
 B3-b1   & -0.78  &   --- &  ---   &  +0.10& --- & ---  & --- & ---  &  +0.30& -0.30 &  -0.15  \\ 
 B3-b2   & 0.18   & 0.00  & -0.30  &  -0.22 &  -0.20 &  +0.15 &  0.00 &  -0.20 &  0.00 & -0.18 & -0.25  \\ 
 B3-b3   & 0.18   & --- & ---&  -0.07  & 0.00 &  +0.10 &  +0.30 &  0.00 &  0.00 &  -0.25 &  -0.05  \\ 
  B3-b4   & 0.17   & 0.05  & -0.30 & -0.12&  +0.15&  -0.05&  -0.20&  -0.10&  +0.25 &  0.00 &  -0.07 \\ 
 B3-b5   & 0.11   & 0.30  & -0.20  & -0.10 & -0.10 & -0.10& 0.00 & 0.00 &  +0.10 & -0.30 &  +0.05 \\ 
 B3-b7   & 0.20   & 0.30  & -0.05  &  0.00 &  +0.25 &  0.00 & 0.00 & -0.15 &  0.00 &  -0.10 &  0.00 \\
 B3-b8   & -0.62  & 0.20  & 0.00   &  0.00  &  0.00 &  0.00 &  0.00 & 0.00 &  0.00 & -0.03 &  0.00  \\ 
 B3-f1   & 0.04   & -0.05&      -0.20 &   -0.10 &  +0.15 & -0.30 & 0.00 &  -0.10 &  +0.20 &  -0.30 &  0.00   \\ 
 B3-f2   & -0.25  & 0.30&       0.30 &  0.00 & -0.20 & +0.25 &  0.00 &  -0.25 &  +0.30 & 0.00 & 0.00   \\ 
 B3-f3   & 0.06   & -0.30&      0.30 &  -0.23 & 0.00 & 0.00 &  +0.15 & -0.15 &  -0.15 & -0.30 &  -0.05   \\ 
 B3-f4   & 0.09   & -0.40&      0.00 &  ---& -0.30 &  0.00 & 0.00 &  0.00 & -0.15 & -0.30 &  +0.05   \\ 
 B3-f5   & 0.16   & -0.40&      -0.40 & -0.15 & 0.00 & -0.30 &  0.00 &  +0.15 & -0.15 & 0.00 & +0.20   \\ 
 B3-f7   & 0.16   & 0.00&       0.00 &  -0.10 & +0.15 & -0.30 & 0.00 &  -0.30 &  0.00 & -0.30 &  +0.25   \\ 
 B3-f8   & 0.20   & 0.50&       0.30 &  -0.10 & -0.10 &  -0.05 &  -0.07 &  -0.20 & +0.10 & 0.00 &  +0.25   \\ 
\hline
\hline
 BWc-1   & 0.09  & 0.00&        0.00 & -0.10 & 0.00 &  0.00 & 0.00 & -0.05 & 0.00 & +0.10 &  0.00 \\
 BWc-2   & 0.18  & -0.60&       -0.60 & -0.30 & 0.00 & -0.15& --- &  -0.10 & -0.30 & -0.30 &  -0.20  \\
 BWc-3  & 0.28   & 0.35&        0.00 & -0.20 & 0.00 & 0.00 & 0.00 & -0.30 & +0.12 & -0.20 &  0.00   \\
 BWc-4  & 0.05   & -0.30&       -0.10 & -0.10 & 0.00 & 0.00 & 0.00 & 0.00 &  0.00 & -0.15 & 0.00  \\ 
 BWc-5  & 0.42   & 0.00&        -0.20 & 0.00 &  0.00 &  0.00 &  0.00 & -0.25 &  +0.15 & -0.20 &  +0.20  \\ 
 BWc-6 &-0.25   & 0.30& -0.30 &  0.00&  0.00 &  0.00 &  0.00& --- &  +0.15 &  -0.08 & 0.00 \\ 
 BWc-7   &-0.25  & -0.30&       0.00  &  0.00 &  +0.10 &  0.00 &  0.00 &  -0.30 &  +0.10 &  -0.25 & ---   \\ 
 BWc-8  &0.37    & 0.00&        -0.10 &  -0.18 & -0.12 &  0.00 &  0.00 &  -0.15 & -0.15 &  -0.20 &  0.00  \\ 
 BWc-9  &0.15    & 0.30&        0.30  &  0.00 & 0.00 & 0.00 & 0.00 &  0.00 &  -0.15 &  -0.15  &  +0.15  \\ 
 BWc-10 &0.07    & -0.30&       -0.30 &   -0.15 &  0.00 &  0.00 & 0.00 &  -0.15 &  -0.10 &  -0.30 &  -0.05 \\
 BWc-11  &0.17   & 0.00&        -0.30 & 0.00 &  0.00 & 0.00 &  0.00 & --- &  -0.05 & --- & -0.30  \\ 
 BWc-12  &0.23   & -0.35&       0.00  &  -0.18 &  0.00 &  -0.05 &  -0.20 &  0.00 &  0.00 &  -0.15 &  0.00  \\
 BWc-13  &0.36   & -0.20&       -0.30 & --- &  -0.05 &  0.00 &  0.00 & --- &  0.00 & --- &  -0.20  \\
\hline
\noalign{\vskip 0.1cm}
\noalign{\hrule\vskip 0.1cm}
\noalign{\vskip 0.1cm}  
\hline                  
\end{tabular}
\end{flushleft}
\end{table*}

\begin{table*}
\begin{flushleft}
\scalefont{0.8}
\caption{Atmospheric parameters and radial velocities adopted from Zoccali et al. (2006) and Lecureur
et al. (2007), and resulting Co and Cu abundances.}             
\label{atmos2}      
\centering          
\begin{tabular}{l@{}c@{}ccccccccccccc}     
\noalign{\smallskip}
\hline\hline    
\noalign{\smallskip}
\noalign{\vskip 0.1cm} 
Star & OGLE no. & $\alpha$(J2000) & $\delta$(J2000)  &
  T$_{\rm eff}$ & logg & [Fe/H] &  v$_{\rm t}$ &
v$_{\rm r}$ & v$_{\rm helio}$ & [Cu/Fe]$\rm_{LTE}$ & [Co/Fe]$\rm_{LTE}$ & [Co/Fe]$\rm_{NLTE}$ \\                            
\noalign{\vskip 0.1cm}
\noalign{\hrule\vskip 0.1cm}
 &  &   &  & [K] &  & & [kms$^{-1}$] & [kms$^{-1}$] & [kms$^{-1}$] \\
\noalign{\vskip 0.1cm}
\noalign{\hrule\vskip 0.1cm}
\noalign{\hrule\vskip 0.1cm}
\noalign{\vskip 0.1cm}  
B6-b1 & 29280c3 & 18 09 50.480  & -31 40 51.61   & 4400 & 1.8 & 0.07 & 1.6 & -88.3 & 11.59  &  -0.20 &      -0.20 &  -0.07 \\
B6-b2 & 83500c6 & 18 10 33.980  & -31 49 09.15   & 4200 & 1.5 &-0.01 &1.4 & 17.0 & 11.66 &     0.03&       -0.15 &   -0.04 \\
B6-b3 & 31220c2 & 18 10 19.060  & -31 40 28.19   & 4700 & 2.0 & 0.10 & 1.6 & -145.8 & 11.64 &  0.03 &        -0.05 & 0.08  \\
B6-b4 & 60208c7 &  18 10 07.770 & -31 52 41.36   & 4400 & 1.9 & -0.41 & 1.7 & -20.3 & 11.61 &  -0.08&        -0.06 &  0.03 \\
B6-b5 & 31090c2 & 18 10 37.380  & -31 40 29.14   & 4600 & 1.9 & -0.37 & 1.3 & -4.2 & 11.67 &   0.13&         -0.05 &  0.06  \\
B6-b6 & 77743c7 & 18 09 49.100  & -31 50 07.66   & 4600 & 1.9 & 0.11 & 1.8 & 44.1 & 11.58 &    -0.03&       -0.08&    0.05  \\
B6-b8 & 108051c7 & 18 09 55.950 & -31 45 46.33   & 4100 & 1.6 & 0.03 & 1.3 & -110.3 & 11.59 &  -0.15&       -0.20&    -0.11  \\

B6-f1 & 23017c3 & 18 10 04.460  & -31 41 45.31   & 4200 & 1.6 & -0.01 & 1.5 & 38.4 & 10.95 &  -0.20 &        -0.14&  -0.03 \\
B6-f2 & 90337c7 & 18 10 11.510  & -31 48 19.28   & 4700 & 1.7 & -0.51 & 1.5 & -98.5 & 10.96 & -0.20 &        -0.08 &  0.05 \\
B6-f3 & 21259c2 & 18 10 17.720  & -31 41 55.20   & 4800 & 1.9 & -0.29 & 1.3 & 90.2 & 10.97 &  0.05 &         +0.01&  0.14 \\
B6-f5 & 33058c2 & 18 10 41.510  & -31 40 11.88   & 4500 & 1.8 & -0.37 & 1.4 & 22.1 & 11.02 &  -0.08 &        -0.04&  0.06 \\
B6-f7 & 100047c6 & 18 10 52.300 & -31 46 42.18   & 4300 & 1.7 & -0.42 & 1.6 & -10.4 & 11.03 & -0.18 &        0.00&   0.09 \\
B6-f8 & 11653c3 & 18 09 56.840  & -31 43 22.56   & 4900 & 1.8 & 0.04 & 1.6 & 58.5 & 10.94 &   0.10 &        -0.01&   0.14 \\

BW-b2 & 214192 & 18 04 23.950 & -30 05 57.80  & 4300 & 1.9 & 0.22 & 1.5 & -19.2 & -6.15 &  -0.15 & -0.11 &  0.00    \\
BW-b4 & 545277 & 18 04 05.340 & -30 05 52.50  & 4300 & 1.4 & 0.07 & 1.4 & 85.6 & -6.18 &   -0.30   & -0.13 & 0.00    \\
BW-b5 & 82760 & 18 04 13.270  & -29 58 17.80   & 4000 & 1.6 & 0.17 & 1.2 & 68.8 & -6.17 &  -0.35   & -0.02 &  0.05    \\
BW-b6 & 392931 & 18 03 51.840 & -30 06 27.90  & 4200 & 1.7 & -0.25 & 1.3 & 140.4 & -6.21 & -0.30   &     -0.06 &  0.04 \\
BW-b7 & 554694 & 18 04 04.570 & -30 02 39.60  & 4200 & 1.4 & 0.10 & 1.2 & -211.1 & -6.19 & -0.25   &      -0.18 &   -0.06  \\

BW-f1 & 433669 & 18 03 37.140 & -29 54 22.30  & 4400 & 1.8 & 0.32 & 1.6 & 202.6 & -2.73 &   -0.40   &     -0.04 &  0.09  \\
BW-f4 & 537070 & 18 04 01.400 & -30 10 20.70  & 4800 & 1.9 &  -1.21 & 1.7 & -144.1 & -2.68 &  -0.80 & 0.00 & 0.22 \\
BW-f5 & 240260 & 18 04 39.620 & -29 55 19.80  & 4800 & 1.9 & -0.59 & 1.3 & -6.1 & -2.61 &     -0.15 &    -0.06 &  0.08  \\
BW-f6 & 392918 & 18 03 36.890 & -30 07 04.30  & 4100 & 1.7 & -0.21 & 1.5 & 182.0 & -2.73 &    -0.50 &     0.00 &  0.08  \\
BW-f7 & 357480 & 18 04 43.920 & -30 03 15.20  & 4400 & 1.9 & 0.11 & 1.7 & -139.5 & -2.60 &    --- & -0.17 &  -0.05  \\
BW-f8 & 244598 & 18 03 30.490 & -30 01 44.80  & 5000 & 2.2 & -1.27 & 1.8 & -24.8 & -2.74 &    -0.65 & +0.05  &  0.35  \\
BL-1 & 1458c3 & 18 34 58.643& -34 33 15.241 & 4500 & 2.1 & -0.16 & 1.5 & 106.6 & -6.37 & 0.05 &   +0.12 & 0.22   \\
BL-3 & 1859c2 & 18 35 27.640& -34 31 59.353 & 4500 & 2.3 & -0.03 & 1.4 & 50.6 & -6.32 & -0.10 &   -0.02  &  0.09  \\
BL-4 & 3328c6 & 18 35 21.240& -34 44 48.217 & 4700 & 2.0 & 0.13 & 1.5 & 117.9 & -6.34 &   0.23 &  +0.00 &  0.14  \\
BL-5 & 1932c2 & 18 36 01.148& -34 31 47.913 & 4500 & 2.1 & 0.16 & 1.6 & 57.9 & -6.27 &  0.00 & -0.08  & 0.05 \\
BL-7 & 6336c7 & 18 35 57.392& -34 38 04.621 & 4700 & 2.4 & -0.47 & 1.4 & 108.1 & -6.27 & 0.08 &   +0.06  & 0.16  \\
B3-b1 & 132160C4 & 18 08 15.840 & -25 42 09.83 & 4300 & 1.7 & -0.78 & 1.5 & -123.8 & 2.32 & --- & -0.01 & 0.07  \\
B3-b2 & 262018C7 & 18 09 14.062 & -25 56 47.35 & 4500 & 2.0 & 0.18 & 1.5 & 7.8 & 2.43 & -0.15   & -0.11 & 0.03  \\
B3-b3 & 90065C3 & 18 08 46.405 & -25 42 44.40  & 4400 & 2.0 & 0.18 & 1.5 & 12.2 & 2.38 &      --- &       0.00 & 0.13 \\
B3-b4 & 215681C6 & 18 08 44.472 & -25 57 56.85 & 4500 & 2.1 & 0.17 & 1.7 & 78.6 & 2.37 &    -0.13&       -0.03 & 0.10 \\
B3-b5 & 286252C7 & 18 09 00.527 & -25 48 06.78 & 4600 & 2 & 0.11 & 1.5 & -51.3 & 2.41 &     0.05& -0.06 & 0.07 \\
B3-b7 & 282804C7 & 18 09 16.540 & -25 49 26.08 & 4400 & 1.9 & 0.20 & 1.3 & 159.7 & 2.44 &   0.13& 0.00& 0.14  \\
B3-b8 & 240083C6 & 18 08 24.602 & -25 48 44.39 & 4400 & 1.8 & -0.62 & 1.4 & -9.6 & 2.34 &   0.10& 0.00 &  0.09  \\
B3-f1 & 129499C4 & 18 08 16.176 & -25 43 19.18 & 4500 & 1.9 & 0.04 & 1.6 & 29.4 & 3.35 &  -0.13& -0.06 &  0.06  \\
B3-f2 & 259922C7 & 18 09 15.609 & -25 57 32.75 & 4600 & 1.9 & -0.25 & 1.8 & 3.4 & 3.46 &  0.30&    +0.01& 0.11 \\
B3-f3 & 95424C3 & 18 08 49.628 & -25 40 36.93  & 4400 & 1.9 & 0.06 & 1.7 & -19.1 & 3.41 & 0.00& -0.09 & 0.03 \\
B3-f4 & 208959C6 & 18 08 44.293 & -26 00 25.05 & 4400 & 2.1 & 0.09 & 1.5 & -81.9 & 3.40 & -0.20  & -0.10 & 0.08 \\
B3-f5 & 49289C2 & 18 09 18.404 & -25 43 37.41  & 4200 & 2.0 & 0.16 & 1.8 & -34.7 & 3.47 &   -0.40& -0.03 & 0.00 \\
B3-f7 & 279577C7 & 18 09 23.694 & -25 50 38.19 &  4800 & 2.1 & 0.16 & 1.7 & -9.2 & 3.48 & 0.00& -0.08 &  0.05  \\
B3-f8 & 193190C5 & 18 08 12.632 & -25 50 04.45 & 4800 & 1.9 & 0.20 & 1.5 & 11.0 & 3.34 &  0.40&   0.00 & 0.15   \\
\hline
\hline
 BWc-1  & 393125   &  18 03 50.445  & -30 05 31.993  & 4476 & 2.1 & 0.09  & 1.5 & ---  & 111.8 &    0.00 &  0.00 & 0.12  \\
 BWc-2  & 545749   &  18 03 56.824  & -30 05 37.390 & 4558 &2.2  &  0.18 & 1.2  & ---  & 62.6 &   -0.60 &  -0.15&  -0.01  \\
 BWc-3  & 564840   &  18 03 54.730  & -30 01 06.096 & 4513 &2.1 &  0.28 & 1.3  & ---  & 237.6 &    0.18 &  -0.07 & 0.08 \\
 BWc-4  & 564857   &  18 03 55.416  & -30 00 57.314 & 4866 &2.2  &  0.05& 1.3  & ---  & 1.1 &     -0.20 &  -0.03 &  0.10 \\ 
 BWc-5  & 575542   &  18 03 56.021  & -29 55 43.716 & 4535 &2.1  &  0.42& 1.5  & ---  & 65.0&    -0.10 &  -0.01& 0.14  \\ 
 BWc-6  & 575585   &  18 03 56.543  & -29 55 11.787 & 4769 &2.2 &  -0.25& 1.3  & ---  & 104.9 &   0.00 &  0.00& 0.11 \\ 
 BWc-7  & 67577    &  18 03 56.543  & -29 55 11.787 & 4590 &2.2 &  -0.25& 1.1  & --- & 0.0 &      -0.15 &  -0.05& 0.05   \\ 
 BWc-8  & 78255    &  18 03 12.494  & -30 03 59.111 & 4610 &2.2  &  0.37& 1.3  & --- & -4.2 &   -0.05 &  -0.10& 0.05  \\ 
 BWc-9  & 78271    &  18 03 16.683  & -30 03 51.406 & 4539 &2.1  &  0.15& 1.5  & --- & 47.8 &      0.30 &         -0.02& 0.11  \\ 
 BWc-10 & 89589    &  18 03 18.914  & -30 01 09.983 & 4793 &2.2  &  0.07& 1.3  & --- &188.0 &    -0.30 &  -0.09& 0.04  \\
 BWc-11 & 89735    &  18 03 04.749  & -29 59 35.301 & 4576 &2.1  &  0.17& 1.0  & --- & 98.0 &    -0.15 &  -0.06& 0.09 \\ 
 BWc-12 & 89832    &  18 03 20.102  & -29 58 25.785 & 4547 &2.1 &  0.23& 1.3  & --- & -47.6 &   -0.18 &  -0.07 &  0.07 \\
 BWc-13 & 89848    &  18 03 04.612  & -29 58 14.080 & 4584 &2.1  &  0.36& 1.1  &  --- & -201.1 &   -0.25 & -0.06&  0.10  \\
\noalign{\vskip 0.1cm}
\noalign{\hrule\vskip 0.1cm}
\noalign{\vskip 0.1cm}  
\hline                  
\end{tabular}
\end{flushleft}
\end{table*}

\begin{table}
\caption{Abundance uncertainties for the metal-rich star B6-f8 and the metal-poor star BW-f8
 for uncertainties of $\Delta$T$_{\rm eff}$ = 100 K,
$\Delta$log g = 0.2,  and $\Delta$v$_{\rm t}$ = 0.2 km s$^{-1}$, and
corresponding total error. The errors are to be
added to reach the reported abundances. } 
\label{errors}
\begin{flushleft}
\begin{tabular}{lcccccccccccc}
\noalign{\smallskip}
\hline
\noalign{\smallskip}
\hline
\noalign{\smallskip}
\hbox{Element} & \hbox{$\Delta$T} & \hbox{$\Delta$log $g$} & 
\phantom{-}\hbox{$\Delta$v$_{t}$} & \phantom{-}\hbox{($\sum$x$^{2}$)$^{1/2}$} &\\
\hbox{} & \hbox{100 K} & \hbox{0.2 dex} & \hbox{0.2 kms$^{-1}$} & & \\
\hbox{(1)} & \hbox{(2)} & \hbox{(3)} & \hbox{(4)} & \hbox{(5)}  &\\
\hline
\hline
\noalign{\smallskip}
\hbox{} & \hbox{} & \hbox{B6-f8} & \hbox{} & \hbox{}  &\\
\noalign{\hrule\vskip 0.1cm}
\hline
\noalign{\smallskip}
\hbox{[CoI/Fe]}      &  +0.09          & +0.01      & +0.01  & 0.09 & \\
\hbox{[CuI/Fe]}      &  +0.12          & +0.02      & +0.00  & 0.12 &  \\
\hline
\noalign{\smallskip}
\hbox{} & \hbox{} & \hbox{BW-f8} & \hbox{} & \hbox{}  &\\
\noalign{\hrule\vskip 0.1cm}
\hline
\noalign{\smallskip}
\hbox{[CoI/Fe]}      &  +0.03          & +0.00      &+0.00   & 0.03 & \\
\hbox{[CuI/Fe]}      &  +0.10          & +0.00      & -0.02  & 0.10 &  \\
\noalign{\smallskip} 
\hline 
\end{tabular}
\end{flushleft}
\end{table}

\section{Chemical evolution models}

We have computed
chemodynamical evolution models for cobalt and copper
for a small classical spheroid with a baryonic mass of 2$\times$10$^9$ M$_{\odot}$
and a dark halo mass $M_{H}$= 1.3$\times$10$^{10}$ M$_{\odot}$,
with the same models presented in
Barbuy et al. (2015) and Fria\c ca \& Barbuy (2017).
The code allows for inflow and outflow of gas, treated
with hydrodynamical equations coupled with chemical evolution.

As decribed in detail in Fria\c ca \& Barbuy (2017),
metallicity dependent yields
from SNe II, SNe Ia, and intermediate mass stars (IMS) are included.
The core-collapse SNII yields  are adopted from WW95.
For lower metallicities we also adopt, in a second calculation,
yields from high explosion-energy hypernovae from 
Nomoto et al. (2013, and references therein).
Yields of SNIa resulting from Chandrasekhar mass white dwarfs
are taken from Iwamoto et al. (1999), namely 
their models W7  (progenitor star of initial metallicity Z=Z$_{\odot}$)
and W70 (initial metallicity Z=0).
The yields for IMS ($0.8 - 8$ M$_{\odot}$)
with initial Z=0.001, 0.004, 0.008, 0.02, and 0.4
are from van den Hoek \& Groenewegen (1997) (variable $\eta_{AGB}$ case).

Specific star formation rates (SFR) are defined as
the inverse of the timescale for the system formation, represented by
$\nu_{\rm SF}$ and  given in Gyr$^{-1}$. It is the ratio of the SFR
 in M$_{\odot}$ Gyr$^{-1}$  over the gas mass in M$_{\odot}$ 
 available  for star formation. In the present models we assume 
 $\nu_{\rm SF}=$ 3 and 1 Gyr$^{-1}$, corresponding to 
  fast timescales  of 0.3  and 1 Gyr, respectively, for the
  chemical enrichment of the bulge.

\begin{figure*}
    \centering
    \includegraphics[width=1.8\columnwidth]{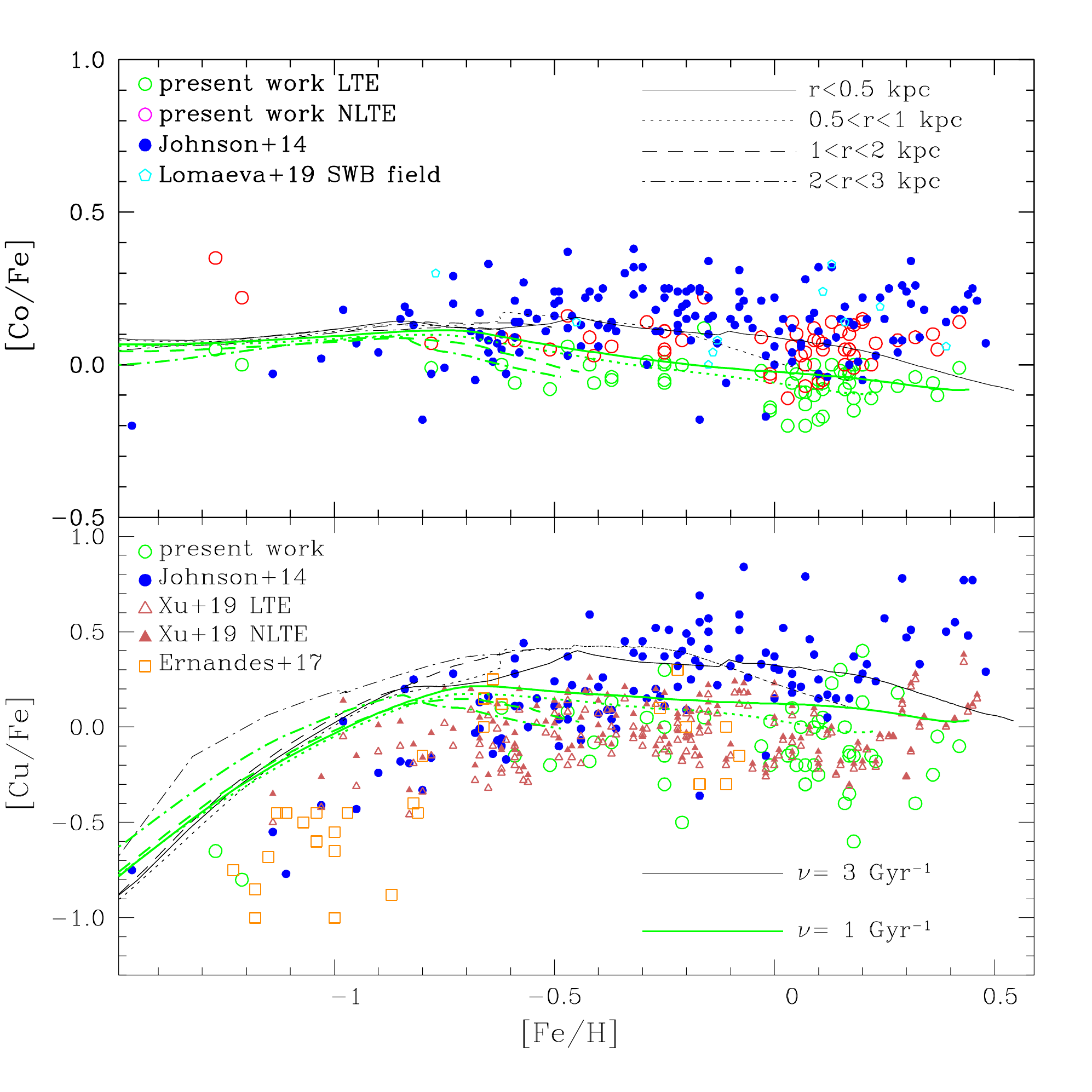}
    \caption{Upper panel: [Co/Fe] vs. [Fe/H] with the present results in LTE and corrected
    for NLTE, together with literature data. Lower panel: [Cu/Fe] vs. [Fe/H] with the present results and literature data.
    Shown are: the present results in LTE (open green circles); present results in NLTE (open magenta circles); Johnson et al. (2014) (filled blue circles); Lomaeva et al. (2019) for the SW field (open cyan circles); Xu et al. (2019) in
    LTE (open Indian red triangles); Xu et al. (2019) in NLTE (filled Indian red triangles); Ernandes et al. (2018) for bulge globular clusters (open dark orange squares) and chemodynamical evolution models
    are overplotted; specific star formation rates of 3 Gyr$^{-1}$,
 with SNII yields from WW95 (black lines); specific star formation rates of 1 Gyr$^{-1}$,
with SNII yields from WW95 and from Kobayashi et al. (2006) for [Fe/H]$<$-4.0 (green lines). Models are for distances to the Galactic center of: r $<$ 0.5 kpc (solid lines), 0.5 $<$
 r $<$ 1 kpc (dotted lines), 1 $<$ r $<$ 2 kpc (dashed lines), and 2 $<$ r
  $<$ 3 kpc (dash-dotted lines).}
    \label{models}
\end{figure*}

 The model calculations overplotted
to the data are shown in Fig. \ref{models}. Models
where only the WW95  yields for massive stars are included are shown in black,
together with a specific star formation rate of 3 Gyr$^{-1}$.
The models in green have a specific star formation rate of 1 Gyr$^{-1}$
and adopting yields from hypernovae
(Kobayashi et al. 2006, Nomoto et al. 2013) instead
of yields from WW95 for metallicities lower than [Fe/H]$<$-4.0.
We have concluded that for these elements (Co, Cu)
the inclusion of hypernovae makes essentially no difference.
Since the yields from core-collapse SNII by  WW95 underestimate
the Co abundance, as recognized by Timmes et al. (1995),
we have multiplied the yields of Co by a factor of two 
for all metallicities Z/Z$_{\odot}$.


In Figure \ref{models}  [Co/Fe] versus [Fe/H] is shown
with the present results in LTE and corrected for NLTE in the upper panel; [Cu/Fe] versus [Fe/H] is shown in the lower panel.
Literature data include:
a) Johnson et al. (2014) and b) Xu et al. (2019), where stars are the same
but they
are plotted as if there were different samples;
c) Lomaeva et al. (2019) only for the stars not in common with the present sample,
which are 11 stars from the SW field (see J\"onsson
et al. 2017). We do not plot the stars in common with the present work
in order to avoid too much clutter in the plot;
d) Ernandes et al. (2018) for bulge globular clusters.

In conclusion, Co is well reproduced by the models, whereas Cu is overproduced.
Chemical evolution  models from Kobayashi et al.
(2006) show a similar Co abundance compatible with the observations,
and also overproduce Cu.

\section{Discussion of results}

Our main interest in the present work is to compare the behaviour
of cobalt and copper. 
They are produced both in the alpha-rich freezeout as
 primary elements (Sukhbold et al. 2016) and in the weak-s
 process in massive stars as secondary elements.
  The iron-peak elements are mainly formed during explosive oxygen
   and silicon burning in massive supernovae (WW95).
   For the larger values of the neutron fraction $\eta$, the main products
   of silicon burning are completed. On the other hand, if the density
   is low and the supernova envelope expansion is fast, $\alpha$ particles
   will be frozen
and not captured by the heavier elements (Woosley et al. 2002).
This so-called $\alpha$-rich freezeout will produce $^{59}$Co.
As pointed out by S. Woosley (private communication) and  
Barbuy et al. (2018a), their abundances as a function of Fe
can reveal the relative efficiencies of these two contributions.

In thick-disc and halo stars, Nissen et al. (2000), 
Cayrel et al. (2004), and Ishigaki et al. (2013),
among others, derived abundances
of iron-peak elements.
Nissen et al. (2000) observed that Sc might be enhanced in
metal-poor stars, and that Mn decreases with decreasing metallicities. Ishigaki et al. (2013) has also shown
that most  Fe-peak elements show solar abundance ratios as a function of metallicity, with the exception of Mn, Cu, and Zn.
In particular as regards Co and Cu, Ishigaki et al. finds that 
Co varies in lockstep with Fe for [Fe/H]$>$-2.0, but appears
enhanced for [Fe/H]$<$-2.0, as previously already found by
Cayrel et al. (2004), and that Cu decreases with
decreasing metallicities.
Barbuy (2013, 2015) and da Silveira et al. (2018) derived Mn and Zn for the present
sample of 56 UVES spectra of red giants, and confirmed that Mn decreases
with decreasing metallicity and that Zn is enhanced in metal-poor stars. 
Ernandes et al. (2018) discussed
 Sc, V, Mn, Cu, and Zn in bulge globular-cluster stars from UVES spectra, with
Sc and V varying in lockstep with Fe, Mn; Cu, increasing
with metallicity; and Zn enhanced in metal-poor stars.
 We will now examine the [Co/Fe] and [Cu/Fe] versus [Fe/H] behaviour.

Before drawing conclusions, we  present literature results
here on Co and Cu in bulge stars.
Johnson et al. (2014) derived abundances of Cr, Co, Ni, and Cu
in 156 giants, 
and Xu et al. (2019) derived Cu abundances for 129 of these same stars,
applying NLTE corrections.
Recently, Lomaeva et al. (2019) derived Sc, V, Cr, Mn, Co, and Ni for bulge giants
that include 33 stars in common using the same UVES data as the present sample.
Schultheis et al. (2017) derived abundances of Cr, Co, Ni, and Mn from APOGEE results,
which show, however, a large spread and are not considered here.


\subsection{Comments on results for cobalt}

Figure \ref{models} shows that [Co/Fe] varies in lockstep with  [Fe/H],
and this appears in all samples.
It appears therefore that the nucleosynthesis process dominating the formation
of cobalt is the alpha-rich freezeout.

\begin{figure*}
    \centering
    \includegraphics[width=1.8\columnwidth]{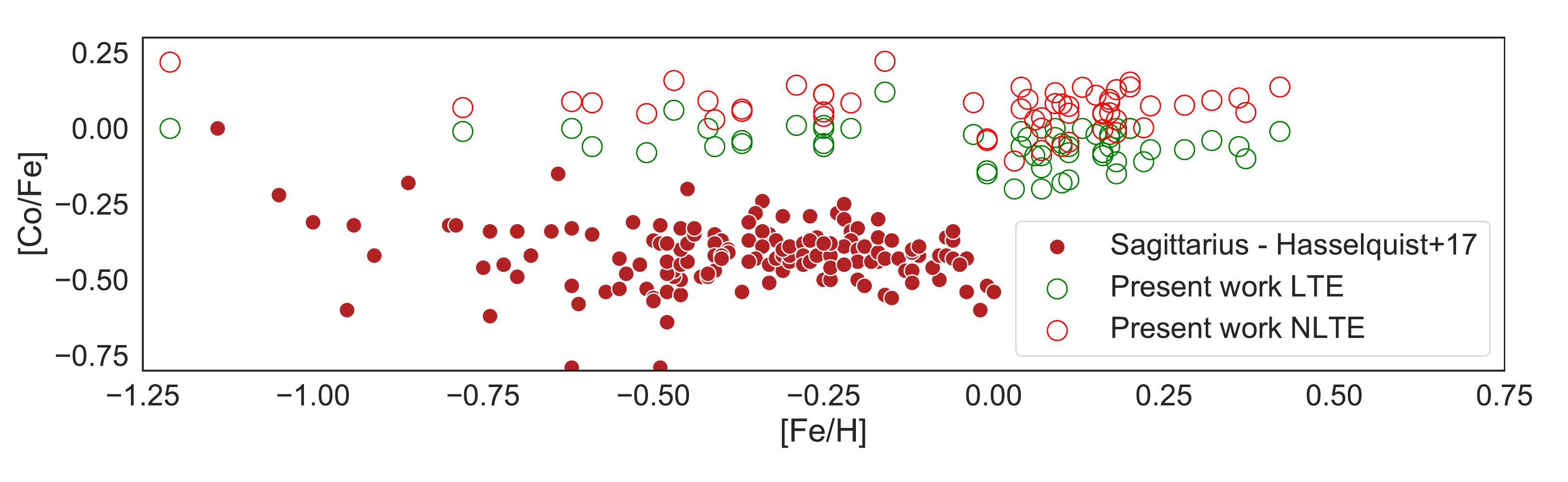}
    \caption{ [Co/Fe] vs. [Fe/H]: present results in LTE and corrected for NLTE,
    compared with data  from Hasselquist et al. (2017) for the Sagittarius 
    dwarf galaxy. 
Symbols: open green circles represent present results in LTE; red circles represent present
results in NLTE; filled blue dots represent Hasselquist et al. (2017).}
    \label{sagitplot}
\end{figure*}

Figure \ref{models} shows that the mean [Co/Fe] value differs among the different authors.
The Johnson et al. (2014) and
Lomaeva et al. (2019) results are in the mean 0.2 dex, more Co-rich than the
present results. A main reason for the discrepancies
  might be the location of continuum.
In order to further investigate   the
  disagreement on  the level of Co deficiency or over-enhancement, it is
interesting to note the deficiency in Co relative to Fe in the Sagittarius
dwarf galaxy. In Fig. \ref{sagitplot} we compare the present results for
Co in LTE and NLTE, compared with Co abundances in 158 red giants of
the Sagittarius dwarf galaxy by Hasselquist et al. (2017). These authors
used the H-band from APOGEE data and found that Co is deficient with
respect to stars in the Milky Way.
Hasselquist et al. (2017) did not consider NLTE effects;
therefore, we compare our results in LTE and theirs, which
leads to a difference in Co abundances of $\Delta$[Co/Fe]$\sim$0.3,
 reduced by 0.2  with respect to results
by Johnson et al. (2014) and Lomaeva et al. (2019).
 Therefore, the deficiency of Co in Sagittarius 
 relative to the present paper is not as drastic as in previous results
 discussed in the literature.
 A possible explanation of the deficiency in Co in Sagittarius, previously
already suggested by McWilliam et al. (2013), is that Sagittarius
was less enriched by SNe II relative to the Milky Way, which could
be caused by a top-light initial mass function (IMF).

\subsection{Comments on results for copper}

In  Fig. \ref{models} all
data agree on [Cu/Fe] versus [Fe/H] having a flat
behaviour between -0.8 $<$ [Fe/H] $<$ +0.1.
For [Fe/H] $<$ -0.8, copper-to-iron clearly
decreases with decreasing metallicity, indicating
the behaviour of a secondary element.
For the metal-rich stars, our data would be compatible
with a flat trend, or a slightly decreasing trend with metallicity,
but this is not shown in the Johnson et al. and
Xu et al. results.
Finally, there is a shift in enhancements between
Johnson et al. and Xu et al. Since they use the same
spectra of the same stars, and the same line,
this could be due to a different placement of continua. Our results fit  the abundance values
from Xu et al. better and we note that the NLTE corrections from Xu et al. are small.

\begin{figure*}
    \centering
    \includegraphics[width=1.8\columnwidth]{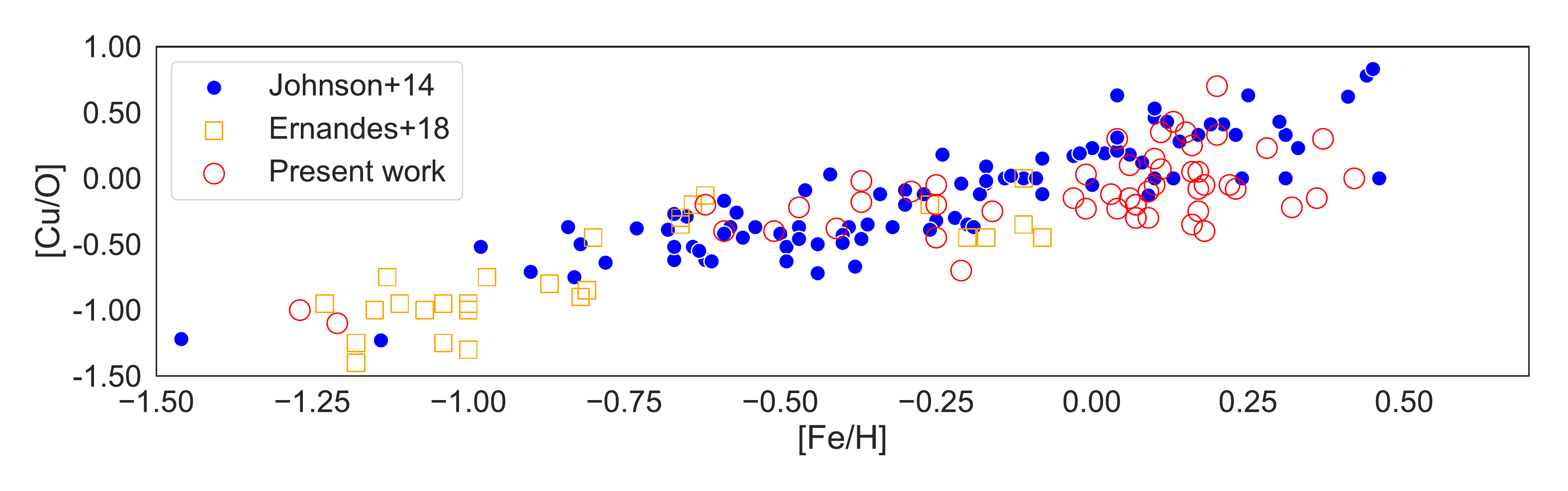}
    \caption{ [Cu/O] vs. [Fe/H] for the present results in NLTE
    and literature data. Symbols: open magenta  circles represent present results in NLTE; filled blue dots represent Jonhson et al. (2014);  open orange squares represent Ernandes et al. (2018).}
    \label{plotox}
\end{figure*}

The behaviour of [Cu/Fe] versus [Fe/H], which shows a decrease in [Cu/Fe] towards
decreasing metallicities, confirms that [Cu/Fe] essentially has
 a secondary-element behaviour
and that its production should be dominated by a weak s-process.
Another characteristic, as noted by McWilliam (2016), is that
[Cu/O] has much less spread than [Cu/Fe] data, indicating a production of Cu and O in the same massive stars.
This is confirmed in Fig. \ref{plotox}, 
where our data are plotted in NLTE together with
data from Johnson et al. (2014) and Ernandes et al. (2018), the latter
corresponding to red giants in bulge globular clusters.
It is clear that the spread of points is lower, confirming the
suggestion by McWilliam (2016).

\section{Conclusions}

We derived the abundances of the iron-peak elements Co and Cu
in 56 red giants of the Galactic bulge, 
for which we have previously derived abundances
of C, N, O, Na, Mg, Al, Mn, Zn, and heavy elements.
The abundances of C, N, O, Na, Mg, Al, Mn, Co, Cu, and Zn are gathered
in Table \ref{tableall}.

Cobalt and copper are in the so-called upper iron-group.
The upper iron-group  elements  Co, Ni, Cu, Zn, Ga, and Ge
with 27 $\leq$ Z $\leq$ 32, or
57$\leq$A$\leq$66 (up to $^{66}$Zn, 
but excluding $^{67,68}$Zn, $^{69}$Ga,$^{70,71}$Ge)
are mainly produced in  two processes, 
namely, a)  neutron capture on iron-group
nuclei during He burning and later burning stages,
also called weak {\it s}-component, 
 and b) the $\alpha$-rich freezeout in the deepest layers
 (Woosley et al. 1973), also discussed in
 WW95, Limongi et al. (2003),  Woosley et al. (2002),
 and Sukhbold et al. (2016). 
 The nucleosynthesis yields from the weak {\it s}-component
 show a characteristic secondary behaviour.
 
 In this work we analysed a sample of  high-quality
 spectroscopic data for 56 Galactic bulge red giants.
 The present results show [Co/Fe] $\sim$ constant $\sim$ 0.0,
 indicating cobalt mainly produced  from the 
 $\alpha$-rich freezeout. Copper instead shows a secondary
 element behaviour, with [Cu/Fe] decreasing with decreasing
 metallicity, indicating its production to be dominated by the
 weak {\it s}-process.
 The yields of Co and Cu considered in the models appear to
 include these two mechanisms in the right proportions, and
 the chemodynamical models reproduce  their behaviour well.

\begin{table*}
\scalefont{0.7}
\caption{Comparison of stellar parameters and Co and Cu abundances of the present work
with Johnson et al. (2014) and Lomaeva et al. (2019). }   
\label{johnson} 
\centering                  
\begin{tabular}{c c c c c c c c c | c c c c c } 
\hline\hline             
 Star &   OGLE  & T$_{\rm eff}$ & logg  & [Fe/H]  & v$_{\rm t}$  & [Co/Fe]$\rm_{LTE}$  & [Co/Fe]$\rm_{NLTE}$  & [Cu/Fe] &  T$_{\rm eff}$  & logg  & [Fe/H]   & [Co/Fe]  &  [Cu/Fe] \\
\hline  
 &   &  & Present work &  &  &  &  &  &   &  & Johnson et al. 2014    &  &   \\
\hline  
B3-b2 &  262018C7 & 4500 &  2.0 &   0.18  &  1.5 &  -0.11 &   0.03 &  -0.15 & 4700 &  2.75 &  0.08 &   0.15 & 0.46 \\
B3-b4 &  215681C6 & 4500 &  2.1 &   0.17  &  1.7 &  -0.03 &   0.10 &  -0.13 & 4800 &  2.75 &  0.31 &   0.20 & 0.51 \\
B3-b5 &  286252C7 & 4600 &  2.0 &   0.11  &  1.5 &  -0.06 &   0.07 &  0.05  & 4700 &  3.10 &  0.43 &   0.18 & 0.77 \\
B3-b7 &  282804C7 & 4400 &  1.9 &   0.20  &  1.3 &   0.00 &   0.14 &  0.13  & 4575 &  2.50 &  0.10 &   0.11 & 0.25 \\
B3-b8 &  240083C6 & 4400 &  1.8 &   -0.62 &  1.4 &   0.00 &   0.09 &  0.10  & 4425 &  1.65 & -0.58 &   0.04 & ---  \\
B3-f1 &  129499C4 & 4500 &  1.9 &   0.04  &  1.6 &  -0.06 &   0.06 &  -0.13 & 4900 &  2.75 &  0.13 &   0.32 & ---  \\
B3-f8 &  193190C5 & 4800 &  1.9 &   0.20  &  1.5 &   0.00 &   0.15 &  0.40  & 4675 &  2.75 &  0.24 &   0.22 & ---  \\
\hline
\hline 
 Star &   OGLE  & Teff & logg  & [Fe/H]  & vt  & [Co/Fe]$\rm_{LTE}$  & [Co/Fe]$\rm_{NLTE}$  & [Cu/Fe] & T$_{\rm eff}$ & logg & [Fe/H]& v$_{\rm t}$& [Co/Fe]$\rm_{LTE}$   \\
\hline
 &   &  & Present work &  &  &  &  &  &   &  & Lomaeva et al. 2019    &  &   \\
\hline
B3-b1& 132160C4 & 4300 & 1.7 & -0.78 & 1.5& -0.01 & 0.07    & ---     &  4414& 1.35& -0.89&  1.41&    0.04    \\
B3-b5& 286252C7 & 4600 & 2.0 & 0.11 & 1.5 & -0.06 & 0.07    & 0.05   & 4425& 2.70& 0.25 &  1.43&   -0.25    \\
B3-b7& 282804C7 & 4400 & 1.9 & 0.20 & 1.3 & 0.00  & 0.14    & 0.13   & 4303& 2.36& 0.08 &  1.58&   -0.08    \\
B3-b8& 240083C6 & 4400 & 1.8 & -0.62 & 1.4& 0.00  &  0.09   & 0.10   &  4287& 1.79& -0.67&  1.46&   0.67    \\
B3-f1& 129499C4 & 4500 & 1.9 & 0.04 & 1.6 & -0.06 &  0.06   & -0.13  &  4485& 2.25& -0.15&  1.88&   0.15    \\
B3-f2& 259922C7 & 4600 & 1.9 & -0.25 & 1.8&  +0.01& 0.11    & 0.30   & 4207& 1.64& -0.66&  1.74&    0.66    \\
B3-f3& 95424C3 & 4400 & 1.9 & 0.06 & 1.7  &  -0.09 & 0.03   & 0.00    & 4637& 2.96& 0.24 &  1.89&    -0.24    \\
B3-f4& 208959C6 & 4400 & 2.1 & 0.09 & 1.5 &  -0.10 & 0.08   & -0.20  & 4319& 2.60& -0.12&  1.50&    0.12    \\
B3-f7& 279577C7 &  4800 & 2.1 & 0.16 & 1.7&  -0.08 &  0.05  & 0.00   &  4517& 2.93& 0.17 &  1.55&  -0.17    \\
B3-f8& 193190C5 & 4800 & 1.9 & 0.20 & 1.5 &  0.00 & 0.15    & 0.40    &  4436& 2.88& 0.24 &  1.54&   -0.24    \\
BW-b1& & & & & & & &                                                  & 4042& 2.39& 0.46 &  1.43&      -0.46    \\
BW-b2& 214192 & 4300 & 1.9 & 0.22 & 1.5   & -0.11  & 0.00   &-0.15   &  4367& 2.39& 0.18 &  1.68&   -0.18    \\
BW-b5& 82760 &  4000 & 1.6 & 0.17 & 1.2   & -0.02  & 0.05   &-0.35     &  3939& 1.68& 0.25 &  1.31&    -0.25    \\
BW-b6& 392931 & 4200 & 1.7 & -0.25 & 1.3   &  -0.06 & 0.04  &-0.30    & 4262& 1.98& -0.32&  1.44&     0.32    \\
BW-b8& & & & & & & &                                                   & 4424& 2.54& 0.30 &  1.52&      -0.30    \\
BW-f1& 433669 & 4400 & 1.8 & 0.32 & 1.6   &  -0.04 &  0.09 & -0.40    &  4359& 2.51& 0.28 &  1.93&   -0.28    \\
BW-f5& 240260 & 4800 & 1.9 & -0.59 & 1.3  &  -0.06 &  0.08 & -0.15    &  4818& 2.89& -0.51&  1.29&    0.51    \\
BW-f6& 392918 & 4100 & 1.7 & -0.21 & 1.5  &  0.00  &  0.08 & -0.50    &  4117& 1.43& -0.43&  1.69&    0.43    \\
B6-b1& 29280c3 & 4400 & 1.8 & 0.07 & 1.6   &-0.20 &  -0.07 & -0.20     &  4372& 2.59& 0.25 &  1.57&  -0.25    \\
B6-b3& 31220c2 & 4700 & 2.0 & 0.10 & 1.6  &      -0.05 &  0.08 &  0.03    &  4468& 2.48& 0.05 &  1.67&  -0.05    \\
B6-b4& 60208c7 & 4400 & 1.9 & -0.41 & 1.7 &      -0.06 &  0.03 &  -0.08   & 4215& 1.38& -0.62&  1.68&    0.62    \\
B6-b5& 31090c2 & 4600 & 1.9 & -0.37 & 1.3 &      -0.05 &  0.06 &  0.13    &  4340& 2.02& -0.48&  1.34&   0.20    \\
B6-b6& 77743c7 & 4600 & 1.9 & 0.11 & 1.8  &      -0.08 &  0.05 & -0.03    &  4396& 2.37& 0.19 &  1.77&   0.23    \\
B6-b8& 108051c7 & 4100 & 1.6 & 0.03 & 1.3 &  -0.20 &  -0.11 & -0.15    &  4021& 1.90& 0.06 & 1.45&   0.06    \\
B6-f1& 23017c3 & 4200 & 1.6 & -0.01 & 1.5 &      -0.14 &  -0.03&  -0.20 &  4149& 2.01& 0.10 &  1.65&   0.10    \\
B6-f3& 21259c2 & 4800 & 1.9 & -0.29 & 1.3 &      +0.01 &  0.14 &  0.05  & 4565& 2.60& -0.35&  1.28&    0.17    \\
B6-f5& 33058c2 & 4500 & 1.8 & -0.37 & 1.4 &      -0.04 &  0.06 &  -0.08 & 4345& 2.32& -0.33&  1.41&    0.29    \\
B6-f7& 100047c6& 4300 & 1.7 & -0.42 & 1.6 & 0.00  &  0.09  & -0.18    & 4250& 2.10& -0.31& 1.65&     0.25    \\
B6-f8& 11653c3 & 4900 & 1.8 & 0.04 & 1.6  &  -0.01 & 0.14  & 0.10   & 4470& 2.78& 0.13 &  1.30&    0.13    \\
BL-1 & 1458c3 & 4500 & 2.1 & -0.16 & 1.5  &      +0.12 & 0.22  & 0.05   &  4370& 2.19& -0.19&  1.50&   0.10    \\
BL-3 & 1859c2 & 4500 & 2.3 & -0.03 & 1.4  &      -0.02 &  0.09 & -0.10  &  4555& 2.48& -0.09&  1.53&   0.16    \\
BL-4 & 3328c6 & 4700 & 2.0 & 0.13 & 1.5   &  +0.00 &  0.14 & 0.23    &  4476& 2.94& 0.27 &  1.41&    0.19    \\
BL-5 & 1932c2 & 4500 & 2.1 & 0.16 & 1.6   &  -0.08 & 0.05  &  0.00   & 4425& 2.65& 0.28 &  1.68&     0.22    \\
BL-7 & 6336c7 & 4700 & 2.4 & -0.47 & 1.4  &      +0.06 & 0.16  & 0.08    & 4776& 2.52& -0.5 &  1.53&    0.19    \\
\hline
\hline
\end{tabular}
\end{table*}

\begin{acknowledgements}
      H.E. acknowledges a PhD fellowship from CAPES (PROEX and PRINT).
      B.B. and A.F. acknowledge partial financial support from the
      brazilian agencies CAPES - Financial code 001, 
      CNPq and FAPESP.
      D.M. and M.Z. gratefully acknowledges support by the BASAL Center for Astrophysics 
and Associated Technologies (CATA) through grant AFB 170002, by the Programa Iniciativa
 Cient\'{\i}fica Milenio grant IC120009, awarded to the Millennium Institute of Astrophysics (MAS),
 and by Proyectos FONDECYT regular No. 1170121 and 1191505.
S.O. acknowledges the partial support of the research program
DOR1901029, 2019, and the project BIRD191235, 2019 of the University of
Padova.
\end{acknowledgements}

\appendix    

\section{Atomic data}

The hyperfine structure constants of \ion{Co}{I} and \ion{Cu}{I} lines employed
in this work are given in Table \ref{coculines}. In Tables
\ref{hfsCo1}, \ref{hfsCo2}, and \ref{hfsCo3}, the lines of \ion{Co}{I} in terms of their HFS components, and corresponding oscillator strengths, are listed.

\begin{table*}
\begin{flushleft}
\scalefont{0.8}
\caption{Atomic constants for CoI and CuI used to compute hyperfine structure:
A and B constants from Pickering (1996) for CoI.  
For CuI, the A and B constants are from Kur\'ucz (1993)
and Biehl (1976), and they are reported in Ernandes et al. (2018).}             
\label{coculines}      
\centering          
\begin{tabular}{lc@{}c@{}c@{}c@{}c@{}c@{}c@{}c@{}c@{}c@{}c@{}c@{}ccc} 
\noalign{\smallskip}
\hline\hline    
\noalign{\smallskip}
\noalign{\vskip 0.1cm} 
Species & $\lambda$ ({\rm \AA}) & \phantom{-}Lower level 
& \phantom{-}J &\phantom{-}A(mK)& \phantom{-}A(MHz) 
&\phantom{-}B(mK)& \phantom{-}B(MHz) & \phantom{-}Upper level  
& \phantom{-}J &\phantom{-}A(mK)& 
\phantom{-}A(MHz) &\phantom{-}B(mK)& \phantom{-}B(MHz)  \\
\noalign{\vskip 0.1cm}
\noalign{\hrule\vskip 0.1cm}
\noalign{\vskip 0.1cm}

$^{59}$CoI & 4749.612 & ($^4$F)4sp z$^6$D &  9/2 &28.05 & 840.9180  & 0.08 & 2.3983 &  ($^5$F)s5s e$^6$F  & 11/2 & 31.45 & 942.8475 & 0.07 & 2.0985  &   \\

$^{59}$CoI & 5212.691 & ($^4$F)4sp z$^4$F  & 9/2 &27.02 & 810.0394  & 0.08 & 2.3983 &  ($^5$F)s5s f$^4$F  & 11/2 & 35.92 &1076.8546 & 0.07 & 2.0985   &   \\

$^{59}$CoI & 5280.629 & ($^4$F)4sp z$^4$G  & 9/2 &17.25 & 517.1420  & 0.09 & 2.6981 &  ($^5$F)s5s f$^4$F  & 7/2  & 28.25 & 846.9138 & 0.07 & 2.0985  &   \\

$^{59}$CoI & 5301.047 & d$^7$s$^4$ a4$^4$P  &  5/2 & 5.90   & 176.8776  & 0.20  & 5.9959 &  ($^3$F)4p y$^4$D  & 5/2  & 15.50  & 464.6784 & 0.20  & 5.9959   &   \\

$^{59}$CoI & 5342.708 & ($^3$F)4p y$^4$G&  11/2 & 10.0 & 299.7925  & 0.20  & 5.9959 &  ($^3$F)4d e$^4$H  & 13/2 & 7.60   & 227.8423 & 0.20  & 5.9959   &   \\

$^{59}$CoI & 5454.572 & ($^3$F)4p y$^4$F  &  9/2 & 9.90 & 296.7946  & 0.10 &  2.9979 &  ($^3$F)4d g$^4$F  & 9/2  & 9.18  & 275.2095 & 0.08 & 2.3983   &   \\

$^{59}$CoI & 5647.234 & ($^3$P)4s a$^2$P &  3/2 &11.20  & 335.7676 & 0.20  & 5.9959  &  ($^3$F)4p y$^3$D  & 5/2  & 16.40  & 491.6597 & 0.10  & 2.9979  &   \\

$^{59}$CoI & 6117.000 & d$^4$s$^2$ a$^4$P &  1/2 &-23.60   & -707.5103 & 0.20 & 5.9959  &  3($^4$F)4sp z$^4$D & 1/2  & 27.50 & 824.4294 & 0.10 & 2.9979 &   \\

$^{59}$CoI & 6188.996 & d$^7$s$^2$ a$^4$P &  5/2 & 5.90    & 176.8776 & 0.08 & 2.3983  &  ($^5$F)4sp z$^4$D  & 5/2  & 23.22 & 696.1182 & 0.09 & 2.6981  &   \\

\noalign{\vskip 0.1cm}
\noalign{\hrule\vskip 0.1cm}
\noalign{\vskip 0.1cm}

$^{63}$CuI & 5105.537 & 4p 2P [case e] & 1.5 &6.5& 194.865 &-0.96& -28.78 &  4s$^2$ $^2$D [case b] & 
2.5  & 24.97& 748.582 &6.20&  185.871   \\ 
    & 5218.197 & 4p 2P [case e]   & 1.5  &6.5& 194.865 &-0.96& -28.78  & 4d $^2$D [---] & 2.5 & 
0.0* & 0.0* & 0.0* & 0.0* \\

\noalign{\vskip 0.1cm}
\noalign{\hrule\vskip 0.1cm}
\noalign{\vskip 0.1cm}

$^{65}$CuI & 5105.537 & 4p 2P [case e] & 1.5 &6.96& 208.66 &-0.86& -25.78 & 4s2 2D [case b] & 
2.5  & 26.79& 803.14 &5.81&  174.18  \\ 
    & 5218.197 & 4p 2P [case e]   & 1.5  &6.96& 208.66 &-0.86& -25.78  & 4d 2D [---] & 2.5 & 
0.0* & 0.0* & 0.0* & 0.0* \\

\noalign{\vskip 0.1cm}
\noalign{\hrule\vskip 0.1cm}
\noalign{\vskip 0.1cm}  
\hline    
              
\end{tabular}
\end{flushleft}
\end{table*}

\begin{table*}
\caption{Hyperfine structure for \ion{Co}{I} lines. }
\label{hfsCo1}
\centering
\begin{tabular}{ccccccccccccc}
\hline
\noalign{\smallskip}
\cline{1-3} \cline{5-7} \cline{9-11} \\
\multicolumn{3}{c}{4749.612\AA;  $\chi$= 3.053 eV} && \multicolumn{3}{c}{5212.691\AA; $\chi$=   3.514 eV}   && \multicolumn{3}{c}{5280.629\AA; $\chi$= 3.629 eV}  & \\
\multicolumn{3}{c}{log gf(total) = $-$0.321} && \multicolumn{3}{c}{log gf(total) = $-$0.110} && \multicolumn{3}{c}{log gf(total) = $-$0.030}   & \\
\noalign{\smallskip}
\cline{1-3} \cline{5-7} \cline{9-11} \\
\noalign{\smallskip}
\cline{1-3} \cline{5-7} \cline{9-11} \\
$\lambda$ (\AA) & log gf & iso && $\lambda$ (\AA) & log gf & iso && $\lambda$ (\AA) & log gf &iso \\
\noalign{\smallskip}
\cline{1-3} \cline{5-7} \cline{9-11} \\
  4746.669 & -1.7470 & 59 &&  5212.691&   -1.5360 & 59 &&   5280.629 &-1.8362 & 59 & \\
  4749.651 & -1.7470 & 59 &&  5212.771&   -1.5360 & 59 &&   5280.660&  -1.8362& 59 & \\
  4749.664 & -2.1985 & 59 &&  5212.786&   -1.9875 & 59 &&   5280.652& -1.7692& 59 & \\
  4749.643 & -1.6287 & 59 &&  5212.757&   -1.4177 & 59 &&   5280.637& -2.4682& 59 & \\
  4749.683 & -3.1985 & 59 &&  5212.808&   -2.9875 & 59 &&   5280.662&  -1.5729& 59 & \\
  4749.662 & -1.9877 & 59 &&  5212.779&   -1.7767 & 59 &&   5280.646&  -1.5729& 59 & \\
  4749.633&  -1.5106 & 59 &&  5212.740&   -1.2996 & 59 &&   5280.623&  -2.3133& 59 & \\
  4749.687&  -2.9925 & 59 &&  5212.808&   -2.7815 & 59 &&   5280.661&  -1.3688& 59 & \\
  4749.659&  -1.8921 & 59 &&  5212.770&   -1.6811 & 59 &&   5280.637&  -1.4682& 59 & \\
  4749.623&  -1.3992 & 59 &&  5212.721&   -1.1882 & 59 &&   5280.605&  -2.3133& 59 & \\
  4749.690&  -2.9645 & 59 &&  5212.806&   -2.7535 & 59 &&   5280.656&  -1.1994& 59 & \\
  4749.655&  -1.8561 & 59 &&  5212.757&   -1.6451 & 59 &&   5280.625&  -1.4212& 59 & \\
  4749.612&  -1.2954 & 59 &&  5212.699&   -1.0844 & 59 &&   5280.585& -2.4102& 59 & \\
  4749.693&  -3.0436 & 59 &&  5212.801&   -2.8326 & 59 &&   5280.649& -1.0532& 59 & \\
  4749.650&  -1.8717 & 59 &&  5212.743&   -1.6607 & 59 &&   5280.609& -1.4268& 59 & \\
  4749.601&  -1.1989 & 59 &&  5212.674&   -0.9879 & 59 &&   5280.563&  -2.6143& 59 & \\
  4749.694&  -3.2355 & 59 &&  5212.794&   -3.0245 & 59 &&   5280.638& -0.9241& 59 & \\
  4749.645&  -1.9539 & 59 &&  5212.726&   -1.7429 & 59 &&   5280.591&  -1.5004& 59 & \\
  4749.588&  -1.1088 & 59 &&  5212.648&   -0.8978 & 59 &&   5280.536& -3.0123& 59 & \\
  4749.695&  -3.6245 & 59 &&  5212.785&   -3.4135 & 59 &&   5280.625&  -0.8082& 59 & \\
  4749.639&  -2.1722 & 59 &&  5212.707&   -1.9612 & 59 &&   5280.570&  -1.7112& 59 & \\
  4749.575&  -1.0245 & 59 &&  5212.619&   -0.8135 & 59 &&   5280.608&  -0.7026& 59 & \\

\noalign{\smallskip}
\hline
\end{tabular}
\end{table*}

\begin{table*}
\caption{Hyperfine structure for \ion{Co}{I} lines. }
\label{hfsCo2}
\centering
\begin{tabular}{ccccccccccccc}
\hline
\noalign{\smallskip}
\cline{1-3} \cline{5-7} \cline{9-11} \\
\multicolumn{3}{c}{5301.047\AA;  $\chi$= 1.710  eV} && \multicolumn{3}{c}{5342.708\AA; $\chi$=   4.021 eV}   && \multicolumn{3}{c}{5454.572\AA; $\chi$=  4.072  eV}  & \\
\multicolumn{3}{c}{log gf(total) = $-$2.000} && \multicolumn{3}{c}{log gf(total) = 0.690} && \multicolumn{3}{c}{log gf(total) = +0.238}   & \\
\noalign{\smallskip}
\cline{1-3} \cline{5-7} \cline{9-11} \\
\noalign{\smallskip}
\cline{1-3} \cline{5-7} \cline{9-11} \\
$\lambda$ (\AA) & log gf & iso && $\lambda$ (\AA) & log gf & iso && $\lambda$ (\AA) & log gf &iso \\
\noalign{\smallskip}
\cline{1-3} \cline{5-7} \cline{9-11} \\
  5301.077&  -3.6513 & 59 &&  5342.700 &  -0.5933 & 59 &&   5454.568&  -1.4018 & 59 & \\
  5301.068&  -3.3960 & 59 &&  5342.708 &  -1.3181 & 59 &&   5454.562&  -1.5981  & 59 & \\
  5301.081 & -3.3960 & 59 &&  5342.700 &  -0.5100 & 59 &&   5454.574 & -1.5981  & 59 & \\
  5301.072&  -5.3045 & 59 &&  5342.720  & -2.5312 & 59 &&   5454.568 & -1.3877  & 59 & \\
  5301.059&  -3.1973 & 59 &&  5342.711  & -1.0998 & 59 &&   5454.560 & -1.3774  & 59 & \\
  5301.077&  -3.1973 & 59 &&  5342.701  & -0.4239 & 59 &&   5454.577 &-1.3774  & 59 & \\
  5301.064&  -4.0615 & 59 &&  5342.726  & -2.2971 & 59 &&   5454.569 & -1.2504  & 59 & \\
  5301.046&  -3.1328 & 59 &&  5342.715  & -1.0058 & 59 &&   5454.558 & -1.2738  & 59 & \\
  5301.070&  -3.1328 & 59 &&  5342.702  & -0.3396 & 59 &&   5454.581&  -1.2738  & 59 & \\
  5301.053&  -3.3442 & 59 &&  5342.732 &  -2.2513 & 59 &&   5454.570& -1.0827  & 59 & \\
  5301.031&  -3.1639 & 59 &&  5342.719 &  -0.9739 & 59 &&   5454.556&  -1.2313  & 59 & \\
  5301.061&  -3.1639 & 59 &&  5342.704 &  -0.2585 & 59 &&   5454.584& -1.2313  & 59 & \\
  5301.040&  -2.9376 & 59 &&  5342.739 &  -2.3183 & 59 &&   5454.571&  -0.9143  & 59 & \\
  5301.013&  -3.3454 & 59 &&  5342.724 &  -0.9941 & 59 &&   5454.554&  -1.2416  & 59 & \\
  5301.049&  -3.3454 & 59 &&  5342.707 &  -0.1810 & 59 &&   5454.588&  -1.2416  & 59 & \\
  5301.023&  -2.6465 & 59 &&  5342.747 &  -2.5012 & 59 &&   5454.572&  -0.7549  & 59 & \\
    &  &                  && 5342.729  &  -1.0809 & 59 &&   5454.553&  -1.3194  & 59 & \\
    &  &                  && 5342.709  &  -0.1073 & 59 &&   5454.593&  -1.3194  & 59 & \\
     &  &                 && 5342.755  &  -2.8833 & 59 &&   5454.573&  -0.6070  & 59 & \\
    &  &                  && 5342.735   & -1.3036 & 59 &&   5454.552&  -1.5340  & 59 & \\
   &  &                   &&  5342.713 &  -0.0370 & 59 &&   5454.597&  -1.5340  & 59 & \\
    &  &                  &&     &  &                  &&   5454.575&  -0.4706  & 59 & \\

\noalign{\smallskip}
\hline
\end{tabular}
\end{table*}

\begin{table*}
\caption{Hyperfine structure for \ion{Co}{I} lines. }
\label{hfsCo3}
\centering
\begin{tabular}{ccccccccccccc}
\hline
\noalign{\smallskip}
\cline{1-3} \cline{5-7} \cline{9-11} \\
\multicolumn{3}{c}{5647.234\AA;  $\chi$= 2.280   eV} && \multicolumn{3}{c}{6117.000\AA; $\chi$= 1.785 eV}   && \multicolumn{3}{c}{6188.996\AA; $\chi$= 1.710  eV}  & \\
\multicolumn{3}{c}{log gf(total) = $-$1.560} && \multicolumn{3}{c}{log gf(total) = $-$2.490} && \multicolumn{3}{c}{log gf(total) = $-$2.450}   & \\
\noalign{\smallskip}
\cline{1-3} \cline{5-7} \cline{9-11} \\
\noalign{\smallskip}
\cline{1-3} \cline{5-7} \cline{9-11} \\
$\lambda$ (\AA) & log gf & iso && $\lambda$ (\AA) & log gf & iso && $\lambda$ (\AA) & log gf &iso \\
\noalign{\smallskip}
\cline{1-3} \cline{5-7} \cline{9-11} \\
 5105.562 & $-$2.8856 & 59 && 5218.195 & $-$1.2041 & 63  && 5218.195 & $-$1.2041 & 59 & \\
  5647.269&  -2.7641  & 59 &&  6117.043&  -3.4511 & 59 &&  6189.071 & -4.1013  & 59 & \\
  5647.258&   -2.7641 & 59 &&  6117.002&  -2.9740 & 59 &&  6189.053 & -3.8460  & 59 & \\
  5647.243&   -3.0652 & 59 &&  6117.008&  -2.9740 & 59 &&  6189.075 &  -3.8460  & 59 & \\
  5647.269&   -2.9402 & 59 &&  6116.967&  -3.1201 & 59 &&  6189.058 &  -5.7545  & 59 & \\
  5647.253&   -2.6003 & 59 &&   &                 &  &&  6189.031 &  -3.6473  & 59 & \\
  5647.232&   -2.6258 & 59 &&   &                 &  &&  6189.064 &  -3.6473  & 59 & \\
  5647.268&   -3.1901 & 59 &&   &                 &  &&  6189.038 &  -4.5115  & 59 & \\
  5647.247&   -2.5924 & 59 &&   &                 &  &&  6189.002 &  -3.5828  & 59 & \\
  5647.220&   -2.3425 & 59 &&   &                 &  &&  6189.046 &  -3.5828  & 59 & \\
  5647.265&   -3.6180 & 59 &&   &                 &  &&  6189.011 &  -3.7942  & 59 & \\
  5647.238&   -2.7527 & 59 &&   &                 &  &&  6188.967 &  -3.6139  & 59 & \\
  5647.207&  -2.1273  & 59 &&   &                 &  &&  6189.022 &  -3.6139  & 59 & \\
          &           &  &&   &                   &  &&  6188.978 &  -3.3876  & 59 & \\
          &           &  &&   &                   &  &&  6188.924 &  -3.7954  & 59 & \\
          &           &  &&   &                   &  &&  6188.991 &  -3.7954  & 59 & \\
          &           &  &&   &                   &  &&  6188.938 &  -3.0965  & 59 & \\

\noalign{\smallskip}
\hline
\end{tabular}
\end{table*}

\section{NLTE corrections to cobalt abundances}

The NLTE corrections to the derived LTE abundances of Co, derived from
calculations made available online by Bergemann et al. (2010) (see text),
are given in Table \ref{atmos4}.

\begin{table*}
\begin{flushleft}
\scalefont{0.7}
\centering
\caption{NLTE corrections to the derived LTE abundances of Co.}             
\label{atmos4}      
\centering          
\begin{tabular}{l@{}cccccccccc}     
\noalign{\smallskip}
\hline\hline    
\noalign{\smallskip}
\noalign{\vskip 0.1cm} 
Star & [Co/Fe]  & [Co/Fe]   & [Co/Fe]  & [Co/Fe]  & [Co/Fe]   & [Co/Fe]  & [Co/Fe]   & [Co/Fe]  & \\ 
     &  5212.691  {\rm \AA}  & 5280.629 {\rm \AA}   & 5301.047 {\rm \AA}   &  5342.708 {\rm \AA}   & 5454.572 {\rm \AA}  & 5647.234 {\rm \AA}   & 6117.000 {\rm \AA}    &  6188.996 {\rm \AA}   &  \\    
\noalign{\vskip 0.1cm}
\noalign{\hrule\vskip 0.1cm}
\noalign{\vskip 0.1cm}  
\noalign{\vskip 0.1cm}
\noalign{\hrule\vskip 0.1cm}
\noalign{\vskip 0.1cm}  

B6-b1  &   0.117  &   0.169 &   0.341 &   0.000 &   0.000 &   0.219 &   0.088 &   0.081 \\ 
B6-b2  &   0.097  &   0.146 &   0.312 &   0.000 &   0.000 &   0.186 &   0.074 &   0.066 \\ 
B6-b3  &   0.124  &   0.185 &   0.320 &   0.000 &   0.000 &   0.216 &   0.094 &   0.106 \\ 
B6-b4  &   0.105  &   0.119 &   0.274 &   0.000 &   0.000 &   0.138 &   0.038 &   0.036 \\ 
B6-b5  &   0.123  &   0.158 &   0.255 &   0.000 &   0.000 &   0.169 &   0.072 &   0.074 \\ 
B6-b6  &   0.122  &   0.181 &   0.333 &   0.000 &   0.000 &   0.220 &   0.088 &   0.097 \\ 
B6-b8  &   0.088  &   0.129 &   0.266 &   0.000 &   0.000 &   0.157 &   0.052 &   0.047 \\ 
B6-f1  &   0.095  &   0.137 &   0.302 &   0.000 &   0.000 &   0.182 &   0.068 &   0.061 \\ 
B6-f2  &   0.158  &   0.199 &   0.245 &   0.000 &   0.000 &   0.192 &   0.123 &   0.115 \\ 
B6-f3  &   0.147  &   0.202 &   0.255 &   0.000 &   0.000 &   0.204 &   0.126 &   0.128 \\ 
B6-f5  &   0.116  &   0.145 &   0.285 &   0.000 &   0.000 &   0.165 &   0.060 &   0.059 \\ 
B6-f7  &   0.101  &   0.116 &   0.302 &   0.000 &   0.000 &   0.139 &   0.036 &   0.031 \\ 
B6-f8  &   0.139  &   0.216 &   0.283 &   0.000 &   0.000 &   0.224 &   0.145 &   0.163 \\ 
BW-b2  &   0.111  &   0.167 &   0.287 &   0.000 &   0.000 &   0.191 &   0.075 &   0.067 \\ 
BW-b4  &   0.111  &   0.186 &   0.342 &   0.000 &   0.000 &   0.214 &   0.102 &   0.091 \\ 
BW-b5  &   0.078  &   0.123 &   0.210 &   0.000 &   0.000 &   0.121 &   0.012 &   0.030 \\ 
BW-b6  &   0.101  &   0.122 &   0.320 &   0.000 &   0.000 &   0.164 &   0.052 &   0.039 \\ 
BW-b7  &   0.105  &   0.177 &   0.322 &   0.000 &   0.000 &   0.196 &   0.087 &   0.077 \\ 
BW-f1  &   0.131  &   0.212 &   0.314 &   0.000 &   0.000 &   0.213 &   0.098 &   0.091 \\ 
BW-f4  &   0.305  &   0.270 &   0.343 &   0.000 &   0.000 &   0.305 &   0.285 &   0.244 \\ 
BW-f5  &   0.188  &   0.220 &   0.243 &   0.000 &   0.000 &   0.212 &   0.155 &   0.140 \\ 
BW-f6  &   0.091  &   0.116 &   0.268 &   0.000 &   0.000 &   0.135 &   0.035 &   0.028 \\ 
BW-f7  &   0.118  &   0.165 &   0.329 &   0.000 &   0.000 &   0.216 &   0.088 &   0.079 \\ 
BW-f8  &   0.407  &   0.347 &   0.468 &   0.000 &   0.000 &   0.412 &   0.415 &   0.363 \\ 
BL-1   &   0.105  &   0.132 &   0.301 &   0.000 &   0.000 &   0.173 &   0.051 &   0.052 \\ 
BL-3   &   0.106  &   0.133 &   0.307 &   0.000 &   0.000 &   0.190 &   0.057 &   0.049 \\ 
BL-4   &   0.128  &   0.194 &   0.329 &   0.000 &   0.000 &   0.227 &   0.100 &   0.109 \\ 
BL-5   &   0.122  &   0.174 &   0.327 &   0.000 &   0.000 &   0.222 &   0.092 &   0.081 \\ 
BL-7   &   0.131  &   0.160 &   0.197 &   0.000 &   0.000 &   0.154 &   0.074 &   0.070 \\ 
B3-b1  &   0.122  &   0.119 &   0.216 &   0.000 &   0.000 &   0.113 &   0.039 &   0.022 \\ 
B3-b2  &   0.131  &   0.193 &   0.348 &   0.000 &   0.000 &   0.233 &   0.106 &   0.092 \\ 
B3-b3  &   0.124  &   0.177 &   0.329 &   0.000 &   0.000 &   0.218 &   0.097 &   0.080 \\ 
B3-b4  &   0.122  &   0.171 &   0.322 &   0.000 &   0.000 &   0.219 &   0.090 &   0.081 \\ 
B3-b5  &   0.125  &   0.185 &   0.340 &   0.000 &   0.000 &   0.229 &   0.091 &   0.092 \\ 
B3-b7  &   0.131  &   0.200 &   0.348 &   0.000 &   0.000 &   0.226 &   0.105 &   0.087 \\ 
B3-b8  &   0.120  &   0.130 &   0.244 &   0.000 &   0.000 &   0.135 &   0.046 &   0.034 \\ 
B3-f1  &   0.115  &   0.168 &   0.339 &   0.000 &   0.000 &   0.216 &   0.080 &   0.079 \\ 
B3-f2  &   0.112  &   0.149 &   0.260 &   0.000 &   0.000 &   0.162 &   0.065 &   0.075 \\ 
B3-f3  &   0.112  &   0.154 &   0.324 &   0.000 &   0.000 &   0.207 &   0.076 &   0.071 \\ 
B3-f4  &   0.108  &   0.148 &   0.304 &   0.000 &   0.000 &   0.200 &   0.074 &   0.063 \\ 
B3-f5  &   0.089  &   0.120 &   0.235 &   0.000 &   0.000 &   0.155 &   0.045 &   0.045 \\ 
B3-f7  &   0.124  &   0.189 &   0.301 &   0.000 &   0.000 &   0.213 &   0.102 &   0.118 \\ 
B3-f8  &   0.141  &   0.222 &   0.342 &   0.000 &   0.000 &   0.248 &   0.130 &   0.141 \\ 
BWc-1  &   0.112  &   0.158 &   0.317 &   0.000 &   0.000 &   0.211 &   0.077 &   0.069 \\ 
BWc-2  &   0.132  &   0.191 &   0.340 &   0.000 &   0.000 &   0.232 &   0.107 &   0.087 \\ 
BWc-3  &   0.146  &   0.221 &   0.355 &   0.000 &   0.000 &   0.239 &   0.117 &   0.095 \\ 
BWc-4  &   0.124  &   0.186 &   0.270 &   0.000 &   0.000 &   0.202 &   0.106 &   0.120 \\ 
BWc-5  &   0.155  &   0.247 &   0.328 &   0.000 &   0.000 &   0.231 &   0.114 &   0.099 \\ 
BWc-6  &   0.125  &   0.174 &   0.231 &   0.000 &   0.000 &   0.178 &   0.090 &   0.095 \\ 
BWc-7  &   0.114  &   0.145 &   0.282 &   0.000 &   0.000 &   0.173 &   0.062 &   0.060 \\ 
BWc-8  &   0.159  &   0.244 &   0.347 &   0.000 &   0.000 &   0.243 &   0.125 &   0.104 \\ 
BWc-9  &   0.125  &   0.178 &   0.331 &   0.000 &   0.000 &   0.226 &   0.093 &   0.084 \\ 
BWc-10 &   0.122  &   0.182 &   0.290 &   0.000 &   0.000 &   0.206 &   0.097 &   0.107 \\ 
BWc-11 &   0.139  &   0.207 &   0.365 &   0.000 &   0.000 &   0.244 &   0.117 &   0.097 \\ 
BWc-12 &   0.141  &   0.209 &   0.353 &   0.000 &   0.000 &   0.240 &   0.116 &   0.097 \\ 
BWc-13 &   0.162  &   0.262 &   0.373 &   0.000 &   0.000 &   0.252 &   0.129 &   0.108 \\ 

\hline
\noalign{\vskip 0.1cm}
\noalign{\hrule\vskip 0.1cm}
\noalign{\vskip 0.1cm}  
\hline                  
\end{tabular}
\end{flushleft}
\end{table*}

\section{Fits of studied lines to the spectra of the Sun, Arcturus,  and $\mu$ Leo}

\begin{figure*}
    \centering
    \subfloat[\ion{Co}{I} 5212.537, 5280.629 {\rm \AA}]{
  \includegraphics[width=85mm]{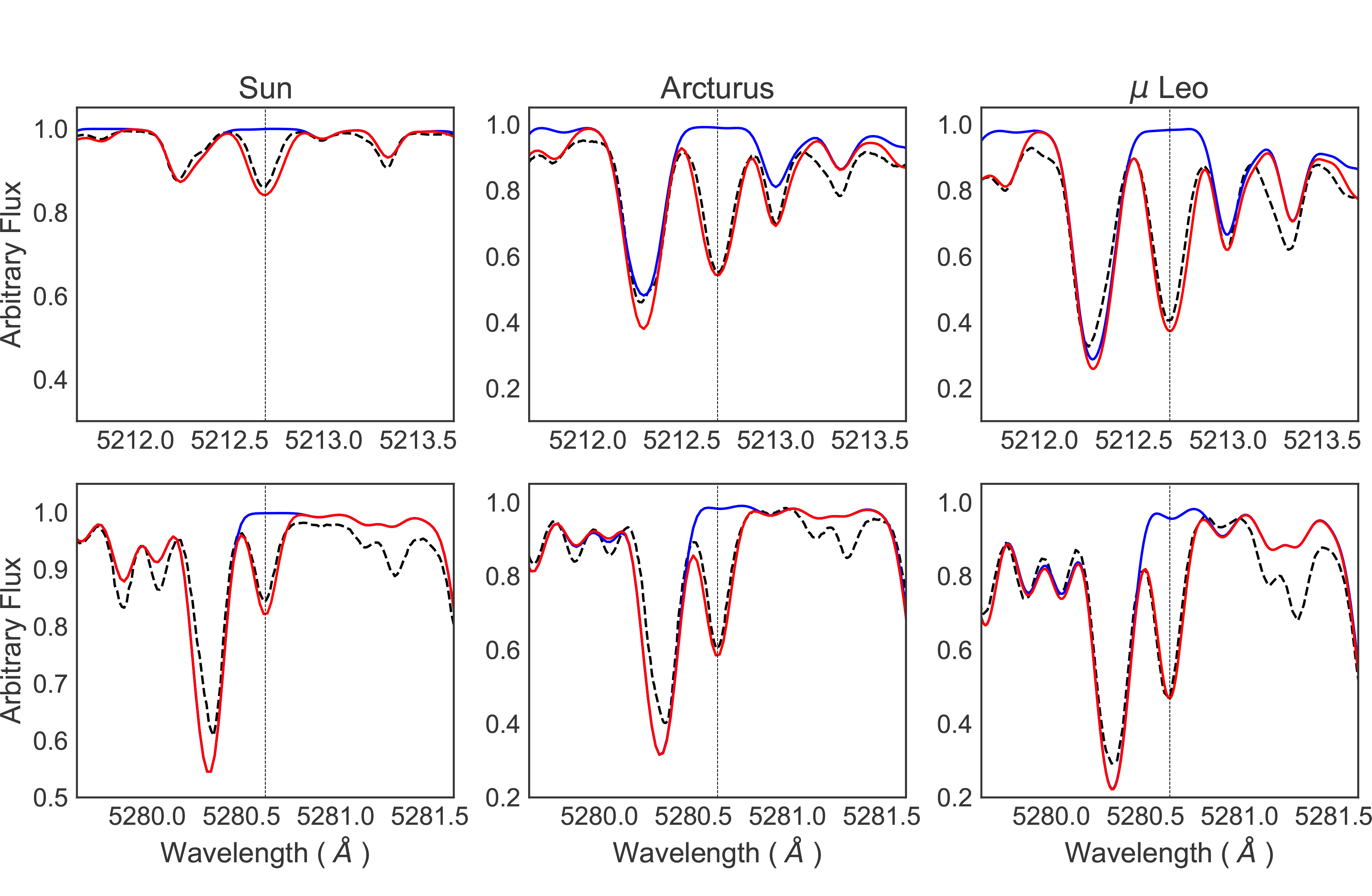}
}
\subfloat[\ion{Co}{I} 5301.039, 5342.695 {\rm \AA}]{
  \includegraphics[width=85mm]{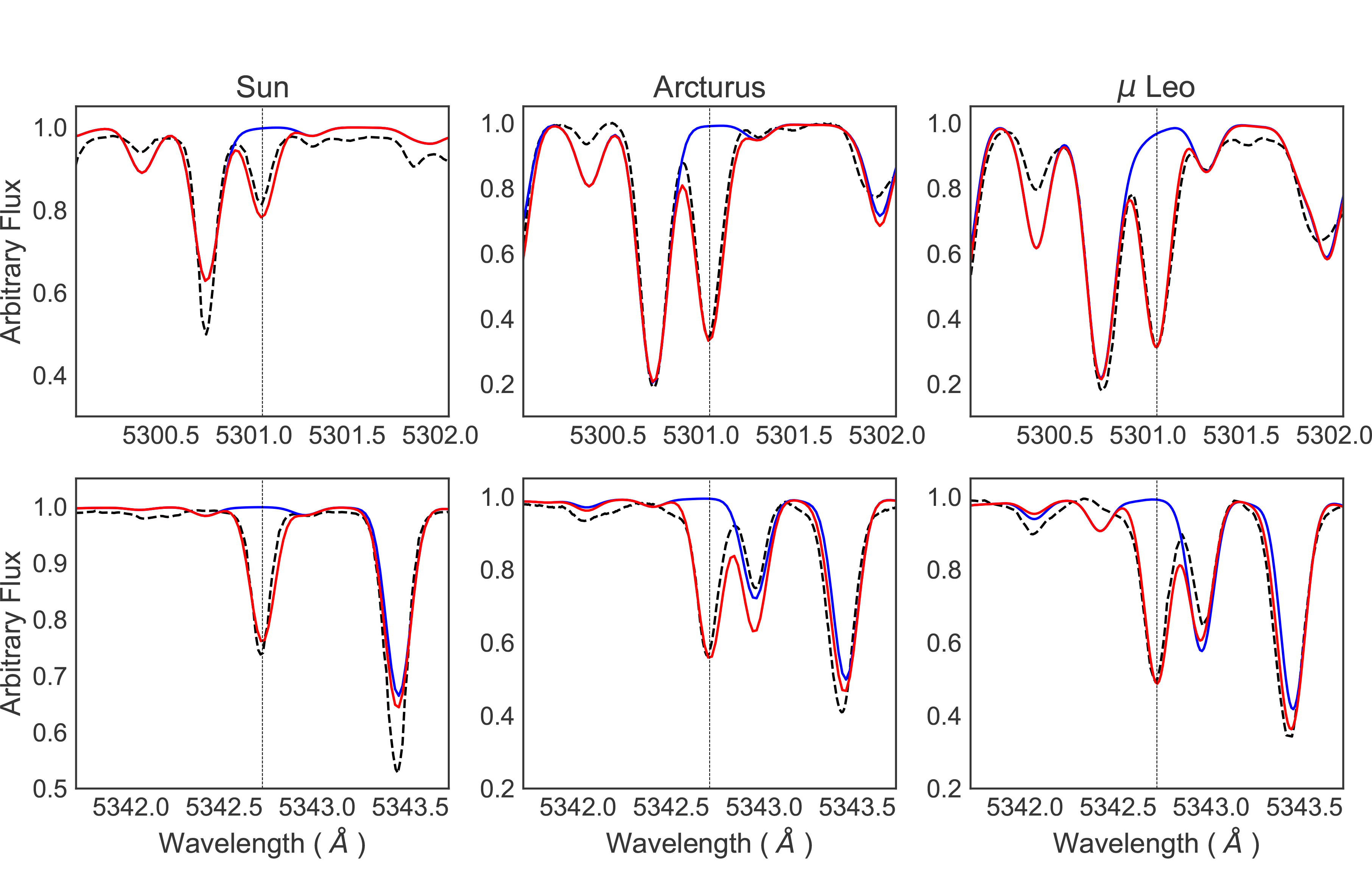}
}
\hspace{0mm}
\subfloat[\ion{Co}{I} 5454.572, 5647.234 {\rm \AA}]{
  \includegraphics[width=85mm]{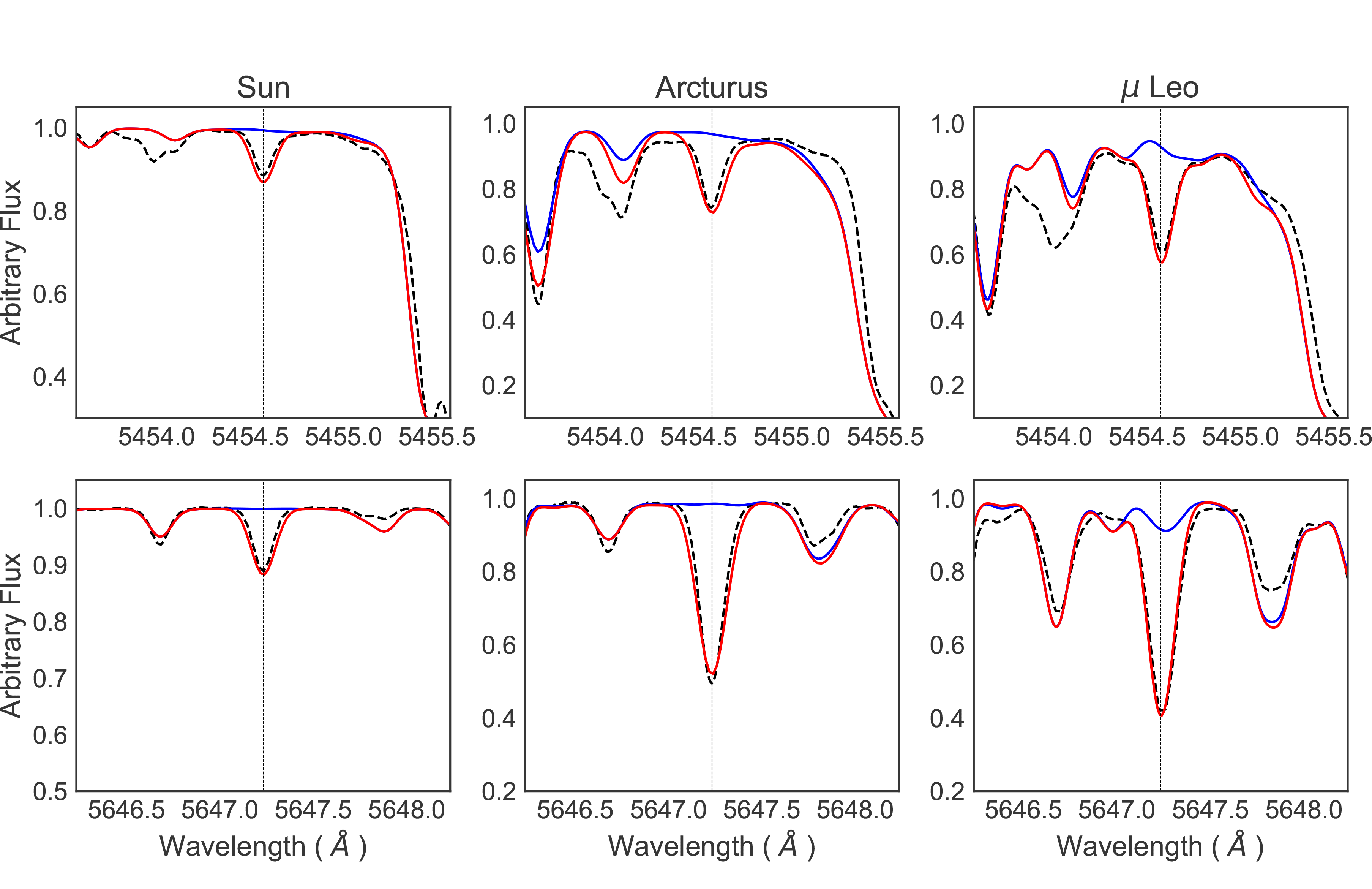}
}
\subfloat[\ion{Co}{I} 6117.0, 6188.996 {\rm \AA}]{
  \includegraphics[width=85mm]{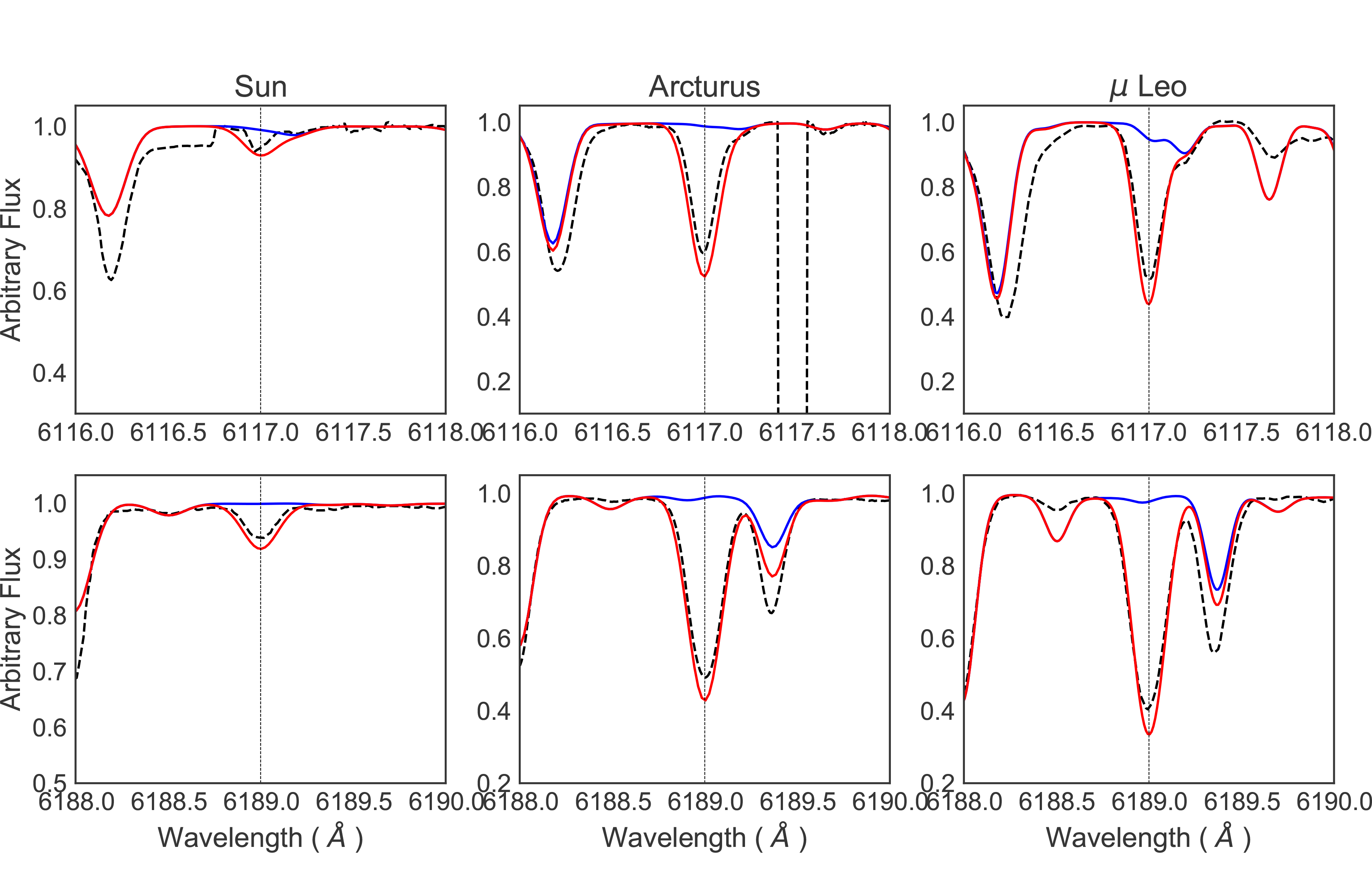}
}
    \caption{Cobalt lines as fitted to the Sun, Arcturus, and $\mu$ Leo.}
    \label{cobaltlines}
\end{figure*}

\begin{figure}
    \centering
    \includegraphics[width=3.5in]{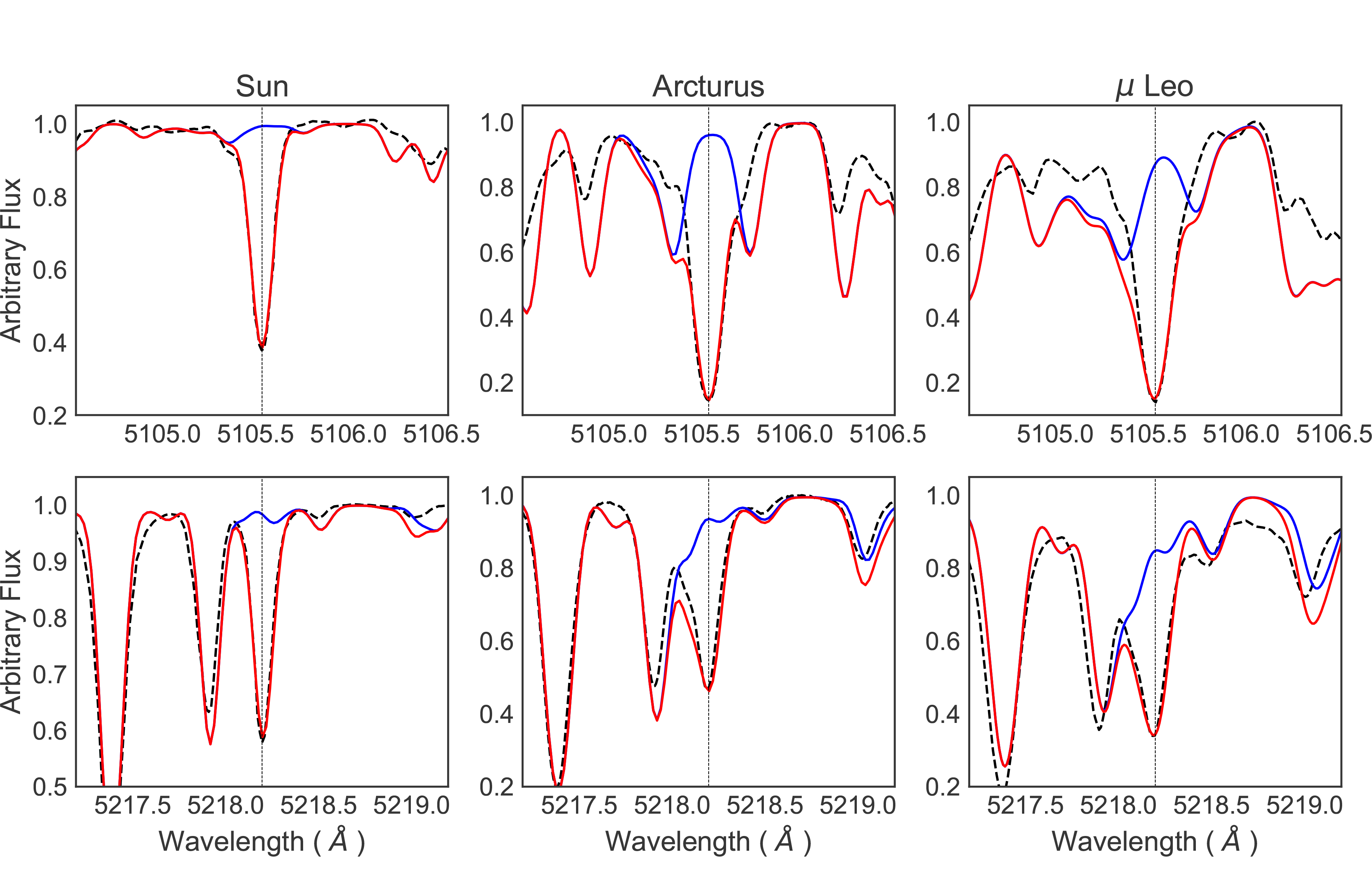}
    \caption{Copper lines \ion{Cu}{I} 5105.537 and 5218.197 {\rm \AA},
    as fitted to the Sun, Arcturus, and $\mu$ Leo.}
    \label{copperlines}
\end{figure}

The lines of \ion{Cu}{I} and \ion{Co}{I} employed to derive abundances in the
present work were first fitted to the spectra of the Sun, Arcturus, and $\mu$ Leo,
as shown in Figs. \ref{cobaltlines} and \ref{copperlines}. Details on the adopted parameters  are given in  Sect. 3.

\section{Abundances of C, N, O, Na, Mg, Al, Mn, Co, and Cu for the 56 sampled red giants}

In Table \ref{tableall}, the 
metallicity from Zoccali et al. (2006), Lecureur
et al. (2007), and Hill et al. (2011) for the 56 sample red giants is reported in column 2 .
The abundances of C, N, O, Na, Mg, Al, Mn, Co, Cu, and Zn from the following
sources are reported:
CNO abundances revised in Fria\c ca \& Barbuy (2017);
Na, Mg, and Al from Lecureur et al. (2007); Mn from Barbuy et al. (2013);
Zn from Barbuy et al. (2015) and da Silveira et al. (2018);
and the present results on Co (LTE and NLTE-corrected) and Cu. 

\begin{table*}
\scalefont{0.9}
\caption{\label{tableall}
Metallicity [Fe/H] and abundances of C, N, O, Na, Mg, Al, Mn, Co, Cu for the 56 sample red giants.} 
\centering                  
\begin{tabular}{lccccccccccccc}
\hline\hline             

Star & [Fe/H] & [C/Fe] & [N/Fe] & [O/Fe] & [Na/Fe] & [Mg/Fe]  & [Al/Fe]  & [Mn/Fe]   &  [Cu/Fe] & [Co/Fe]$\rm_{LTE}$  & [Co/Fe]$\rm_{NLTE}$  & [Zn/Fe]  \\

\hline  
B6-b1   &    0.07 &   -0.15 &    0.50 &    0.00 &    0.57 &    0.21  &    0.59 &     0.06 &   -0.20 &   -0.20 &   -0.07&  -0.20  \\
B6-b2   &   -0.01 &   -0.05 &    0.35 &    0.00 &       --- &   ---  &     --- &    -0.03 &    0.03 &   -0.15 &   -0.04&  -0.15  \\ 
B6-b3   &    0.10 &   -0.25 &    0.50 &   -0.12 &    0.45 &    0.21  &    0.43 &     0.00 &    0.03 &   -0.05 &    0.08&  -0.27  \\
B6-b4   &   -0.41 &   -0.15 &    0.15 &    0.30 &    0.15 &    0.41  &    0.31 &    -0.20 &   -0.08 &   -0.06 &    0.03&   0.00  \\
B6-b5   &   -0.37 &   -0.10 &    0.30 &    0.15 &    0.32 &    0.42  &    0.58 &    -0.04 &    0.13 &   -0.05 &    0.06&   0.10  \\
B6-b6   &    0.11 &   -0.15 &    0.50 &   -0.10 &    0.68 &    0.32  &    0.67 &     0.00 &   -0.03 &   -0.08 &    0.05&  -0.40  \\
B6-b8   &    0.03 &    0.00 &    0.10 &   -0.03 &    0.46 &    0.31  &    0.50 &    -0.03 &   -0.15 &   -0.20 &   -0.11&  -0.08  \\
B6-f1   &   -0.01 &    0.05 &    0.20 &    0.03 &    0.20 &    0.24  &    0.41 &     0.02 &   -0.20 &   -0.14 &   -0.03&  -0.30  \\
B6-f2   &   -0.51 &    0.00 &    0.20 &    0.20 &    0.22 &    0.44  &    0.57 &    -0.08 &   -0.20 &   -0.08 &    0.05&   0.05  \\
B6-f3   &   -0.29 &   -0.05 &    0.30 &    0.15 &    0.31 &    0.43  &    0.53 &     0.00 &    0.05 &    0.01 &    0.14&   0.10  \\
B6-f5   &   -0.37 &    0.05 &    0.00 &    0.10 &    0.23 &    0.41  &    0.74 &    -0.08 &   -0.08 &   -0.04 &    0.06&   0.10  \\
B6-f7   &   -0.42 &    0.00 &    0.30 &   --- &      0.22 &    0.54  &    0.68 &     0.00 &   -0.18 &    0.00 &    0.09&  -0.15  \\ 
B6-f8   &    0.04 &   -0.10 &    0.30 &   -0.20 &    0.50 &    0.27  &    0.72 &     0.00 &    0.10 &   -0.01 &    0.14&  -0.60  \\
BW-b2   &    0.22 &   -0.10 &    0.20 &   -0.10 &    0.01 &    0.40  &    0.26 &     0.00 &   -0.15 &   -0.11 &    0.00&  -0.15  \\
BW-b4   &    0.07 &   -0.10 &    0.00 &   -0.10 &     --- &     ---  &     --- &     0.00 &   -0.30 &   -0.13 &    0.00&   0.00  \\ 
BW-b5   &    0.17 &    0.00 &    0.05 &   -0.10 &    0.37 &    0.19  &    0.49 &     0.00 &   -0.35 &   -0.02 &    0.05&  -0.30  \\
BW-b6   &   -0.25 &    0.00 &    0.65 &    0.15 &    0.22 &    0.59  &    0.55 &     0.00 &   -0.30 &   -0.06 &    0.04&   0.00  \\
BW-b7   &    0.10 &   -0.25 &    0.10 &   -0.20 &     --- &     ---  &     --- &     0.00 &   -0.25 &   -0.18 &   -0.06&  -0.30  \\ 
BW-f1   &    0.32 &   -0.20 &    0.45 &   -0.18 &    0.93 &    0.46  &    0.49 &     0.00 &   -0.40 &   -0.04 &    0.09&  -0.35  \\
BW-f4   &   -1.21 &    0.30 &    0.30 &    0.30 &   -0.06 &    0.42  &    0.86 &    -0.72 &   -0.80 &    0.00 &    0.22&   0.30  \\
BW-f5   &   -0.59 &    0.10 &    0.40 &    0.25 &    0.23 &    0.45  &    0.50 &     0.00 &   -0.15 &   -0.06 &    0.08&   0.15  \\
BW-f6   &   -0.21 &    0.08 &    0.40 &    0.20 &   -0.08 &    0.61  &    0.25 &     0.00 &   -0.50 &    0.00 &    0.08&   0.15  \\
BW-f7   &    0.11 &   -0.20 &    0.70 &   -0.25 &    0.36 &    0.29  &    0.26 &     0.00 &     --- &   -0.17 &   -0.05&  -0.20  \\ 
BW-f8   &   -1.27 &    0.00 &    0.20 &    0.35 &    9.99 &    0.56  &    9.99 &    -0.60 &   -0.65 &    0.05 &    0.35&   0.30  \\
BL-1    &   -0.16 &    0.15 &    0.40 &    0.30 &    0.17 &    0.32  &    0.44 &    -0.01 &    0.05 &    0.12 &    0.22&   0.05  \\ 
BL-3    &   -0.03 &    0.07 &    0.00 &    0.05 &    0.03 &    0.35  &    0.40 &    -0.02 &   -0.10 &   -0.02 &    0.09&   0.10  \\ 
BL-4    &    0.13 &   -0.10 &    0.20 &   -0.20 &    0.70 &    0.39  &    0.78 &     0.00 &    0.23 &    0.00 &    0.14&  -0.30  \\ 
BL-5    &    0.16 &    0.00 &    0.40 &   -0.05 &    0.51 &    0.32  &    0.58 &     0.00 &    0.00 &   -0.08 &    0.05&  -0.27  \\ 
BL-7    &   -0.47 &    0.00 &    0.30 &    0.30 &    0.06 &    0.46  &    0.36 &    -0.30 &    0.08 &    0.06 &    0.16&   0.30  \\ 
B3-b1   &   -0.78 &    0.00 &    0.60 &    0.35 &    0.04 &    0.53  &    0.40 &    -0.35 &     --- &   -0.01 &    0.07&   0.30  \\ 
B3-b2   &    0.18 &     --- &    0.20 &   -0.10 &    0.27 &    0.35  &    0.19 &     0.00 &   -0.15 &   -0.11 &    0.03&  -0.10  \\
B3-b3   &    0.18 &   -0.10 &    0.00 &   -0.20 &    0.46 &    0.37  &    0.37 &     0.00 &     --- &    0.00 &    0.13&    ---  \\ 
B3-b4   &    0.17 &   -0.15 &    0.40 &   -0.05 &    0.49 &    0.50  &    0.27 &     0.00 &   -0.13 &   -0.03 &    0.10&   0.00  \\
B3-b5   &    0.11 &   -0.20 &    0.00 &   -0.30 &    0.56 &    0.32  &    0.59 &     0.00 &    0.05 &   -0.06 &    0.07&   0.00  \\
B3-b7   &    0.20 &   -0.15 &    0.25 &   -0.20 &    0.34 &    0.12  &    0.39 &     0.00 &    0.13 &    0.00 &    0.14&  -0.50  \\
B3-b8   &   -0.62 &   -0.15 &    0.15 &    0.30 &   -0.02 &    0.47  &    0.34 &    -0.10 &    0.10 &    0.00 &    0.09&   0.30  \\
B3-f1   &    0.04 &    0.00 &    0.40 &    0.10 &    0.45 &    0.35  &    0.52 &     0.00 &   -0.13 &   -0.06 &    0.06&   0.00  \\
B3-f2   &   -0.25 &     --- &     --- &     --- &    0.53 &    0.55  &    0.66 &     0.00 &    0.30 &    0.01 &    0.11&   0.00  \\ 
B3-f3   &    0.06 &    0.00 &    0.00 &   -0.10 &    0.34 &    0.54  &    0.25 &     0.00 &    0.00 &   -0.09 &    0.03&   0.00  \\
B3-f4   &    0.09 &    0.00 &    0.10 &    0.10 &    9.99 &    0.20  &    9.99 &     0.00 &   -0.20 &   -0.03 &    0.08&   0.03  \\
B3-f5   &    0.16 &   -0.05 &    0.50 &   -0.05 &     --- &     ---  &     --- &     0.00 &   -0.40 &   -0.09 &   -0.00&   0.15  \\ 
B3-f7   &    0.16 &    0.00 &    0.20 &   -0.25 &     --- &     ---  &     --- &     0.00 &    0.00 &   -0.08 &    0.05&    ---  \\ 
B3-f8   &    0.20 &   -0.20 &    0.30 &   -0.30 &     --- &     ---  &     --- &     0.20 &    0.40 &    0.00 &    0.15&  -0.60  \\ 
\hline
\hline
BWc-1  &    0.09 &    0.05 &    0.30 &    0.10 &    0.24 &    0.29  &    0.41 &     0.05 &   0.00 &    0.00 &    0.12&  -0.45  \\ 
BWc-2  &    0.18 &   -0.20 &    0.15 &   -0.20 &    0.13 &    0.21  &    0.35 &    -0.16 &  -0.60 &   -0.15 &   -0.01&  -0.20  \\ 
BWc-3  &    0.28 &   -0.10 &    0.40 &   -0.05 &    0.54 &    0.12  &    0.56 &     0.06 &   0.18 &   -0.07 &    0.08&    ---  \\ 
BWc-4  &    0.06 &   -0.10 &    0.05 &   -0.05 &    0.10 &    0.44  &    0.52 &     0.03 &  -0.20 &   -0.03 &    0.10&  -0.30  \\ 
BWc-5  &    0.42 &   -0.05 &    0.30 &   -0.10 &    0.72 &    0.01  &    0.60 &     0.20 &  -0.10 &   -0.01 &    0.14&  -0.35  \\ 
BWc-6  &   -0.25 &   -0.20 &    0.70 &    0.05 &    0.22 &    0.52  &    0.47 &    -0.04 &   0.00 &    0.00 &    0.11&   0.00  \\ 
BWc-7  &   -0.25 &   -0.20 &    0.30 &    0.05 &    0.21 &    0.39  &    0.26 &    -0.30 &  -0.15 &   -0.05 &    0.05&   0.00  \\ 
BWc-8  &    0.37 &   -0.30 &    0.10 &   -0.35 &    0.23 &    0.21  &    0.17 &    -0.06 &  -0.05 &   -0.10 &    0.05&   0.00  \\ 
BWc-9  &    0.15 &   -0.10 &    0.20 &   -0.05 &    0.14 &    0.08  &    0.32 &     0.05 &   0.30 &   -0.02 &    0.11&  -0.05  \\ 
BWc-10 &    0.07 &   -0.20 &    0.30 &    0.00 &    0.11 &    0.31  &    0.41 &    -0.06 &  -0.30 &   -0.09 &    0.04&   0.00  \\ 
BWc-11 &    0.17 &   -0.20 &    0.00 &   -0.20 &    0.29 &    0.18  &    0.31 &    -0.10 &  -0.15 &   -0.06 &    0.09&  -0.05  \\ 
BWc-12 &    0.23 &   -0.15 &    0.05 &   -0.10 &    0.49 &    0.30  &    0.59 &     0.10 &  -0.18 &   -0.07 &    0.07&  -0.45  \\ 
BWc-13 &    0.36 &    0.00 &   -0.15 &   -0.10 &   -0.03 &    0.27  &    0.35 &    -0.05 &  -0.25 &   -0.06 &    0.10&  -0.20  \\ 

\hline                          
\end{tabular}
\end{table*}

\end{document}